%% file: ms.tex
\documentclass[11pt]{article}
\pdfoutput=1 %
\usepackage[sc]{mathpazo}

\usepackage{palatino,microtype}
\usepackage[margin=15pt,font=small,labelfont={bf}]{caption}
\usepackage[margin=1in]{geometry}
\usepackage[english]{babel}
\usepackage[utf8]{inputenc}%
\usepackage[compact]{titlesec}
\usepackage{cmap}
\usepackage[T1]{fontenc}
\usepackage{bm}
\usepackage{booktabs}
\usepackage{mathtools}
\usepackage{amsfonts}
\usepackage{amsmath}
\usepackage{amssymb}
\usepackage{amsthm}
\usepackage{float}
\usepackage{graphics}
\usepackage{natbib}
\usepackage[backref=page]{hyperref}
\usepackage[svgnames]{xcolor}
\hypersetup{colorlinks={true},urlcolor={blue},linkcolor={DarkBlue},citecolor=[named]{DarkGreen},linktoc=all}
\renewcommand*{\backref}[1]{}
\renewcommand*{\backrefalt}[4]{%
    \ifcase #1 (Not cited.)%
    \or        (Cited on page~#2)%
    \else      (Cited on pages~#2)%
    \fi}

\usepackage{microtype}
\usepackage{doi}

\usepackage[ruled, vlined, linesnumbered]{algorithm2e}

\SetCommentSty{mycommfont}

\usepackage[capitalise,nameinlink,noabbrev]{cleveref}
\usepackage{standalone}
\usepackage{tikz}
\usepackage{multirow}
\usepackage{nicefrac}
\usepackage{colortbl}
\usepackage{xcolor}
\colorlet{mycolor}{blue!15}
\usepackage{thmtools,thm-restate}
\usepackage{enumitem}
\usepackage{etoolbox}

\newcommand{\BibTeX}{\rm B\kern-.05em{\sc i\kern-.025em b}\kern-.08em\TeX}

\SetKwProg{Init}{initialize}{}{}
\SetKwFunction{pEnvyFreeCycle}{EnvyFreeCycle}
\SetKwFunction{pLocalSearch}{LocalSearch}
\SetKwFunction{pFairOrEfficient}{MakeFairOrEfficient}
\SetKwFunction{pEnvyFreeCycle}{EnvyFreeCycle}
\SetKwFunction{pLocalSearch}{LocalSearch}
\SetKwFunction{pFairOrEfficient}{MakeFairOrEfficient}

\newtheorem{theorem}{Theorem}[section]
\newtheorem{lemma}[theorem]{Lemma}

\newtheorem*{claim*}{Claim}

\newtheorem{example}{Example}

\theoremstyle{remark}
\newtheorem*{remark}{Remark}

\theoremstyle{definition}

\crefname{theorem}{Theorem}{Theorems}
\crefname{lemma}{Lemma}{Lemmas}
\crefname{proposition}{Proposition}{Propositions}
\crefname{corollary}{Corollary}{Corollaries}
\crefname{fact}{Fact}{Facts}
\crefname{definition}{Definition}{Definitions}
\crefname{remark}{Remark}{Remarks}
\crefname{section}{Section}{Sections}
\crefname{appendix}{Appendix}{Appendices}
\crefname{algorithm}{Algorithm}{Algorithms}

\newcommand{\RR}{{\mathbb R}}

\newcommand{\QN}{{\mathbb Q}_{\ge0}}
\newcommand{\ZN}{{\mathbb Z}_{\ge0}}
\newcommand{\ZP}{{\mathbb Z}_{>0}}
\newcommand{\cR}{\mathcal{R}}

\newcommand{\cT}{\mathcal{T}}

\newcommand{\I}{{\mathcal I}}
\newcommand{\W}{{\mathcal W}}
\newcommand{\V}{{\mathcal V}}
\newcommand{\cost}{\text{cost}}
\renewcommand{\O}{{\mathcal O}}
\renewcommand{\P}{\textup{\textsf{P}}}
\newcommand{\PTAS}{\textup{\textsf{PTAS}}}
\newcommand{\QPTAS}{\textup{\textsf{QPTAS}}}
\newcommand{\NP}{\textup{\textsf{NP}}}
\newcommand{\NPh}{\textup{\textsf{NP-hard}}}

\newcommand{\poly}{\textup{poly}}
\newcommand{\APXh}{\textup{\textsf{APX-hard}}}

\newcommand{\eps}{\varepsilon}

\DeclareMathOperator*{\NSW}{NSW} 
\DeclareMathOperator{\OPT}{\texttt{OPT}} 
\DeclareMathOperator{\ALG}{\texttt{ALG}} 
\DeclareMathOperator*{\argmax}{arg\,max}

\newif\ifcomments
\commentstrue
\ifcomments
\newcommand{\SG}[1]{{\textcolor{orange}{SG: { #1}}}}
\else
\newcommand{\SG}[1]{}
\fi

\newif\ifcomments
\commentstrue
\ifcomments
\newcommand{\HS}[1]{{\textcolor{blue}{HS: { #1}}}}
\else
\newcommand{\HS}[1]{}
\fi

\newif\ifcomments
\commentstrue
\ifcomments
\newcommand{\RV}[1]{{\textcolor{magenta}{RV: { #1}}}}
\newcommand{\JY}[1]{{\textcolor{purple}{JY: { #1}}}}
\else
\newcommand{\RV}[1]{}
\newcommand{\JY}[1]{}
\fi

\usepackage{tabularx,tabulary,booktabs,array}
\usepackage{framed}

\newcommand{\problembox}[3]{
\begin{center}
	{\small 
		\begin{tabularx}{1.0\columnwidth}{ll}
			\toprule
			\multicolumn{2}{c}{#1} \\
			\midrule
			\textbf{Input:}& \parbox[t]{0.8\columnwidth}{#2
				\vspace*{1mm}} \\%
			\textbf{Goal:}& \parbox[t]{0.8\columnwidth}{#3\vspace*{1mm}} \\ 
			\bottomrule
		\end{tabularx}
	}
\end{center}
\smallskip
}

\newcommand{\OneSidedNash}{\textup{\textsc{Uncapacitated One-Sided NSW}}}
\newcommand{\ShufflingSwap}{\textup{\textsc{Full Swap}}}
\newcommand{\AlteringSwap}{\textup{\textsc{Partial Swap}}}

\newcommand{\CapacitatedOneSidedNash}{\textup{\textsc{Capacitated One-Sided NSW}}}
\newcommand{\TwoSidedNash}{\textup{\textsc{Uncapacitated Two-Sided NSW}}}
\newcommand{\CapacitatedTwoSidedNash}{\textup{\textsc{Capacitated Two-Sided NSW}}}
\newcommand{\MinCostFLow}{\textup{\textsc{Min-Cost-Flow}}}

\newcommand{\ExactCapacitatedOneSidedNash}{\textup{\textsc{Exact Capacitated One-Sided NSW}}}

\tikzstyle{tikzfig}=[baseline=-0.25em,scale=0.5]
\pgfdeclarelayer{nodelayer}
\pgfdeclarelayer{edgelayer}
\pgfsetlayers{edgelayer,nodelayer,main}
\tikzstyle{std_node}=[fill=white, draw=black, shape=circle]
\tikzstyle{none}=[inner sep=0mm]

\title{Approximating One-Sided and Two-Sided\\Nash Social Welfare With Capacities}
\author{
	\begin{tabular}{m{0.14\linewidth}m{0.14\linewidth}m{0.14\linewidth}m{0.14\linewidth}m{0.14\linewidth}m{0.14\linewidth}}
		\multicolumn{2}{c}{\textbf{Salil Gokhale}} & \multicolumn{2}{c}{\textbf{Harshul Sagar}} & \multicolumn{2}{c}{\textbf{Rohit Vaish}}\\
		\multicolumn{2}{c}{\small{IIT Delhi, India}} & \multicolumn{2}{c}{\small{IIT Delhi, India}} & \multicolumn{2}{c}{\small{IIT Delhi, India}}\\
		\multicolumn{2}{c}{\href{mailto:mt1210237@iitd.ac.in}{\small{\texttt{mt1210237@iitd.ac.in}}}} & \multicolumn{2}{c}{\href{mailto:ee1210211@ee.iitd.ac.in}{\small{\texttt{ee1210211@ee.iitd.ac.in}}}} & \multicolumn{2}{c}{\href{mailto:rvaish@iitd.ac.in}{\small{\texttt{rvaish@iitd.ac.in}}}}\\
        &&&\\
        \multicolumn{3}{c}{\textbf{Vignesh Viswanathan}} & \multicolumn{3}{c}{\textbf{Jatin Yadav}}\\
		\multicolumn{3}{c}{\small{University of Massachusetts, Amherst, USA}} & \multicolumn{3}{c}{\small{IIT Delhi, India}}\\
		\multicolumn{3}{c}{\href{mailto:vviswanathan@umass.edu}{\small{\texttt{vviswanathan@umass.edu}}}} & \multicolumn{3}{c}{\href{jatin.yadav@cse.iitd.ac.in}{\small{\texttt{jatin.yadav@cse.iitd.ac.in}}}}\\
	\end{tabular}
}

\date{}

\begin{document}

\maketitle

\begin{abstract}
We study the problem of maximizing \emph{Nash social welfare}, which is the geometric mean of agents' utilities, in two well-known models. The first model involves \emph{one-sided} preferences, where a set of indivisible items is allocated among a group of agents (commonly studied in fair division). The second model deals with \emph{two-sided} preferences, where a set of workers and firms, each having numerical valuations for the other side, are matched with each other~(commonly studied in matching-under-preferences literature). We study these models under \emph{capacity constraints}, which restrict the number of items (respectively, workers) that an agent (respectively, a firm) can receive. 

We develop constant-factor approximation algorithms for both problems under a broad class of valuations. Specifically, our main results are the following: (a) For any $\eps > 0$, a $(6+\eps)$-approximation algorithm for the one-sided problem when agents have \emph{submodular} valuations, and (b) a $1.33$-approximation algorithm for the two-sided problem when the firms have \emph{subadditive} valuations. The former result provides the first constant-factor approximation algorithm for Nash welfare in the one-sided problem with submodular valuations and capacities, while the latter result improves upon an existing $\sqrt{\texttt{OPT}}$-approximation algorithm for additive valuations. Our result for the two-sided setting also establishes a computational separation between the Nash and utilitarian welfare objectives. We also complement our algorithms with hardness-of-approximation results. Additionally, for the case of additive valuations, we modify the configuration LP of ~\cite{feng_et_al:LIPIcs.ICALP.2024.63} to obtain an $(e^{1/e}+\epsilon)-$ approximation algorithm for {\it weighted} two-sided Nash social welfare under capacity constraints.
\end{abstract}

\section{Introduction}
\label{sec:Introduction}

Fairness and efficiency are quintessential requirements in many resource allocation problems, such as distributing a set of items among agents and assigning job applicants to positions. These objectives are often represented as opposite ends of the ``collective utility'' scale~\citep{M04fair}: On one end, there is Bentham's \emph{utilitarian} welfare, which maximizes the sum of individual utilities and represents a purely efficient outcome. On the other end, there is Rawls' \emph{egalitarian} welfare (and its leximin refinement), which captures perfect fairness by maximizing the worst-off individual's utility. 

The \emph{Nash social welfare}~\citep{N50bargaining,KN79nash}, defined as the geometric mean of agents' utilities, strikes a remarkable balance between fairness and efficiency. It provides a ``sweet spot'' between these seemingly incompatible objectives and satisfies several desirable properties such as scale invariance, Pigou-Dalton principle, (approximate) envy-freeness, and Pareto optimality~\citep{M04fair,CKM+19unreasonable}. Due to its strong axiomatic appeal, the computational aspects of Nash welfare have received significant attention~\citep{EG59consensus,NNR+14computational,L17apx,CG18approximating,BKV18finding,CKM+19unreasonable,ACH+22maximizing,GKK23approximating,GHL+23approximating,JV24maximizing}.

In this work, we focus on the problem of maximizing Nash social welfare in two well-studied models of resource allocation:
\begin{itemize}%
    \item \emph{One-sided} preferences: In this model, a set of indivisible items is allocated among a set of agents. Each agent has combinatorial preferences over the items, while each item can be assigned to exactly one agent. This setting is commonly explored in the literature on fair division with indivisible items~\citep{BCE+16handbook,M17approximation,AAB+23fair}.
    \item \emph{Two-sided} preferences: In this model, a set of workers is matched with a set of firms. Each firm can be matched with multiple workers and has combinatorial preferences over them. Each worker has cardinal valuations over the firms and can only be assigned to one firm. This model is frequently studied in the matching-under-preferences literature~\citep{GI89stable,RS92two,K97stable,M13algorithmics,BCE+16handbook}.
\end{itemize}

An important feature distinguishing our work from much of the prior work is the consideration of \emph{capacity} constraints. In the context of the one-sided problem, this means restricting the number of items that an agent can receive, while in the two-sided setting, each firm is restricted to be matched with at most a certain number of workers. We allow different agents to have different capacities.

Capacity constraints naturally capture some of the practical limitations %
in resource allocation problems. For example, when distributing pieces of artwork among museums, the space limitation of each museum restricts the number of items it can accommodate~\citep{S21constraints}. Similarly, the hiring capacity of a firm is often limited by its financial budget. In such scenarios, it is natural to aim for outcomes that maximize welfare while adhering to capacity constraints.

It is known that maximizing Nash welfare is hard to approximate (specifically, it is \APXh{}) in the one-sided problem even when there are no capacity constraints~\citep{NNR+14computational,L17apx}. In the two-sided model, the problem of maximizing Nash welfare is known to be \NPh{} even when each firm has a constant capacity~\citep{JV24maximizing}. Since the uncapacitated setting is a special case of the capacitated model, the latter problem is computationally more challenging. Therefore, it is important to develop approximation algorithms for these problems.

\begin{table*}[t]
    \centering
    \begin{tabular}{@{\extracolsep{1.5pt}}c c c c c c@{}}\\
    \multirow{2}{*}{\textbf{One-sided}} & \multicolumn{2}{c}{Nash} & \multicolumn{2}{c}{Utilitarian}\\
    \cline{2-3}
    \cline{4-5}
    & Hardness & Algorithm & Hardness & Algorithm\\
    \midrule
    Without & $\frac{e}{e-1}-\eps$ & $4+\eps$ & $\frac{e}{e-1}-\eps$ & $\frac{e}{e-1}$ \\
    Capacities & %
    \citep{GKK23approximating} & \citep{GHL+23approximating} & \citep{KLM+08inapproximability} & \citep{V08optimal} \vspace{0.2cm} \\
    With & $\frac{e}{e-1}-\eps$ & \cellcolor{mycolor} $6+\eps$ & $\frac{e}{e-1}-\eps$ & $\frac{e}{e-1}$ \\
    Capacities & %
    \citep{GKK23approximating} & \cellcolor{mycolor} (\Cref{theorem:capacitated-1-sided}) & \citep{KLM+08inapproximability} & \citep{V08optimal} \\
    \end{tabular}
    \begin{tabular}{@{\extracolsep{1.5pt}}c c c c c c@{}}\\
    \multirow{2}{*}{\textbf{Two-sided}} & \multicolumn{2}{c}{Nash} & \multicolumn{2}{c}{Utilitarian}\\
    \cline{2-3}
    \cline{4-5}
    & Hardness & Algorithm & Hardness & Algorithm\\
    \midrule
    Without & \cellcolor{mycolor}1.0000759 & \cellcolor{mycolor}1.33 (subadditive) & $\frac{e}{e-1}-\eps$  & $\frac{e}{e-1}$  \\
    Capacities & \cellcolor{mycolor}(\Cref{thm:Two_Sided_Hardness_Nash}) & \cellcolor{mycolor}(\Cref{theorem:capacitated-2-sided}) & \citep{KLM+08inapproximability} & \citep{V08optimal} \vspace{0.2cm} \\
    With & \cellcolor{mycolor}1.0000759 & \cellcolor{mycolor}1.33 (subadditive) & $\frac{e}{e-1}-\eps$ & $\frac{e}{e-1}$\\
    Capacities & \cellcolor{mycolor}(\Cref{thm:Two_Sided_Hardness_Nash}) & \cellcolor{mycolor}(\Cref{theorem:capacitated-2-sided}) & \citep{KLM+08inapproximability} & \citep{V08optimal}%
    \vspace{1mm}
    \end{tabular}
    \caption{Summary of results on maximizing Nash social welfare for the one-sided (top) and two-sided  (bottom) problems under \emph{submodular} valuations. The rows specify whether or not capacities are considered. The columns specify the best-known approximation algorithms and hardness results~(assuming $\P{} \neq \NP{}$) for Nash and utilitarian welfare. Our contributions are highlighted in shaded boxes. Note that our $1.33$-approximation algorithm for the two-sided problem applies to the more general domain of \emph{subadditive} valuations. %
    }
    \label{tab:Summary}
\end{table*}

\subsection*{Our Contributions}
We present constant-factor approximation algorithms for maximizing Nash welfare in the one-sided and two-sided models under capacity constraints for a broad class of valuations. Our results are summarized below (also see \Cref{tab:Summary}). Throughout, we will use $\eps>0$ to denote an arbitrary constant.
\begin{itemize}%
    \item \emph{One-sided model}: In \Cref{sec:OneSidedAlgo}, we provide a $(6+\eps)$-approximation algorithm for maximizing Nash welfare under capacity constraints when the agents have monotone \emph{submodular} valuations over the items. Our algorithm runs in strongly polynomial time. In the same setting without capacity constraints, a local search-based $(4+\eps)$-approximation algorithm was recently proposed by \citet{GHL+23approximating}. Although their algorithm does not automatically handle capacity constraints, we show that with some necessary modifications (such as allowing two-way exchange of items instead of only one-way transfers), the algorithm can be adapted to the more general problem with capacities, albeit with a slight loss in approximation quality. On the hardness front, \citet{GKK23approximating} have shown that, unless $\P{}=\NP{}$, no algorithm that makes a polynomial number of value queries can provide a better approximation of Nash welfare than $\frac{e}{e-1} \approx 1.58$. This intractability holds even for a constant number of agents with submodular valuations and even without capacity constraints and, therefore, extends to the capacitated setting.   
    \item \emph{Two-sided model}: In \Cref{sec:TwoSidedAlgo}, we present a $1.33$-approximation strongly-polynomial time algorithm for maximizing Nash welfare when the firms have capacity constraints and monotone \emph{subadditive} valuations over the workers, and the workers have cardinal valuations over individual firms. Prior to our work, a $\sqrt{\texttt{OPT}}$-approximation for positive additive valuations was known, where $\texttt{OPT}$ denotes the optimal Nash welfare~\citep{JV24maximizing}. Thus, our result presents a significant improvement over the current algorithm in both approximation quality and the generality of preferences. Notably, our result also establishes that in the two-sided setting, Nash welfare is computationally \emph{easier} than utilitarian welfare for which a hardness result of $\frac{e}{e-1} - \eps$ is known~\citep{KLM+08inapproximability}. Our algorithm and its analysis are simple and use a single minimum-cost flow computation. We also show that the same algorithm can be used to provide a \PTAS{} when the number of firms is constant. 
    \item \emph{Hardness results}: In \Cref{sec:HardnessResults}, we show that maximizing Nash welfare in the two-sided problem is \APXh{}; specifically, it is \NPh{} to approximate Nash welfare within a factor of $1.0000759$ even without capacity constraints. This observation strengthens an existing intractability result which only shows \NPh{}ness~\citep{JV24maximizing}. In light of the aforementioned \PTAS{}, it follows that our \APXh{}ness result cannot be extended to the case of a constant number of firms unless $\P{}=\NP{}$.
\end{itemize}

\subsection*{Related Work}
\label{subsec:Related-Work}

We will now review some of the relevant literature on one-sided and two-sided Nash welfare. %

\paragraph{One-sided preferences.} 
In the one-sided model without capacity constraints, it is known that maximizing Nash social welfare is \APXh{} even under additive valuations~\citep{NNR+14computational,L17apx,GHM24satiation,FVZ24hardness}. A substantial body of work has studied the design of exact and approximation algorithms for Nash welfare over various classes of valuations, including \emph{additive}~\citep{CG18approximating,CDG+17convex,BKV18finding,BKV18greedy,ACH+22maximizing}, \emph{budget-additive}~\citep{GHM18approximating}, \emph{separable piecewise-linear concave}~\citep{AMG+18nash,CCG+22fair}, \emph{matroid rank}~\citep{VZ23general,BCI+21finding,BEF21fair}, \emph{submodular}~\citep{GKK23approximating,GHL+23approximating}, and \emph{subadditive}~\citep{BBK+20tight,CGM21fair,DLR+24constant}. These valuation classes have been extensively studied in algorithmic game theory~\citep{NRT+07agt} and computational social choice~\citep{BCE+16handbook}. In particular, additive valuations are attractive from an elicitation perspective, while submodular valuations capture the idea of diminishing marginal returns.

For submodular valuations,~\citet{GHL+23approximating} have shown a $(4+\eps)$-approximation algorithm in the \emph{value query} model (see \Cref{sec:Preliminaries} for the definition). In the same setting, it is known that any algorithm that makes a polynomial number of value queries fails to provide better than $\frac{e}{e-1} \approx 1.58$ approximation~\citep{GKK23approximating}. (This result uses the reduction of~\citet{KLM+08inapproximability} on the hardness of approximating \emph{utilitarian} welfare under submodular valuations.) For the case of a constant number of agents, \citet{GKK23approximating} provide an $\frac{e}{e-1}$-approximation algorithm, matching the hardness threshold.

For subadditive valuations, an $\O(n)$ approximation algorithm is known for value queries~\citep{BBK+20tight,CGM21fair}. This bound is essentially tight due to a hardness result of $\Omega(n^{1-\eps})$~\citep{BBK+20tight}. In the more powerful model with \emph{demand} queries, a constant-factor ($\approx 375,000$) approximation was recently shown by \citet{DLR+24constant} for subadditive valuations.\footnote{In the demand query, the input consists of a set of prices $p_1,p_2,\dots,p_m$ of the items and an index $i$, and the response consists of a bundle $S$ that maximizes $v_i(S) - \sum_{j \in S} p_j$. A value query can be simulated by a polynomial number of demand queries. However, simulating a demand query may require an exponential number of value queries~\cite[Chapter 11]{NRT+07agt}.}

\paragraph{One-sided model with capacity constraints.}

Several papers have studied fair division under cardinality constraints, and, more generally, matroid constraints~\citep{GM14near,BB19matroid,WLG21budget,DFS23fair,SHS23efficient,GNC+23towards}. However, with the exception of~\citep{WLG21budget} and~\citep{GNC+23towards}, these works are primarily focused on fairness notions such as approximate envy-freeness, equitability, and proportionality instead of algorithms for maximizing Nash welfare. \citet{WLG21budget} show that in the presence of budget constraints~(of which capacity constraints are a special case), a maximum Nash welfare allocation may not satisfy envy-freeness up to one item (EF1); however, an $\alpha$-approximate Nash welfare allocation is $1/4\alpha$-EF1, where $\alpha > 1$.\footnote{An allocation is said to be $\alpha$-envy-free up to one item ($\alpha$-EF1) if for every pair of agents $i,k$, there exists some item $g \in A_k$ such that $v_i(A_i) \geq \frac{1}{\alpha} v_i(A_k \setminus \{g\})$~\citep{B11combinatorial}. In the presence of capacity constraints, $\alpha$-EF1 requires that for any pair of agents $i,k$ and any subset $S \subseteq A_k$ such that $|S| \leq c_i$, there exists some item $g \in S$ such that $v_i(A_i) \geq \frac{1}{\alpha} v_i(S \setminus \{g\})$~\citep{WLG21budget}.}~\citet{GNC+23towards} study a model where each item has a fixed number of copies that can be assigned to different agents. Their model involves one-sided preferences and two-sided cardinality constraints, i.e., a lower bound on the number of items assigned to an agent and an upper bound on the number of agents receiving copies of an item. We refer the reader to the survey by~\citet{S21constraints} for a detailed review of the literature on constraints in fair division.

It is relevant to note that an instance with additive valuations and capacity constraints can be converted to an unconstrained instance with submodular valuations. Indeed, for any set $S$ of items, we define agent $i$'s value $v_i(S)$ as the sum of values of the $\min\{c_i,|S|\}$ most valuable items in $S$, where $c_i$ is the capacity for agent $i$. The resulting valuation function $v_i$ can be observed to be submodular. Thus, an $\alpha$-approximation algorithm for the resulting submodular valuations constitutes an $\alpha$-approximation for the original additive valuations instance with capacity constraints. Consequently, for any additive valuations instance with capacity constraints, the algorithm of \citet{GHL+23approximating} provides a $(4+\eps)$ approximation to Nash welfare. However, when the given instance has submodular valuations with capacity constraints, such a transformation may not result in an unconstrained submodular instance; see \Cref{sec:OneSidedAlgo} for an example. Thus, one can no longer use existing approximation algorithms for unconstrained submodular valuations to solve the constrained problem.

\paragraph{Two-sided preferences.} The two-sided matching problem has been extensively studied in economics, computer science, and artificial intelligence~\citep{GI89stable,RS92two,K97stable,M13algorithmics,BCE+16handbook} and has several real-world applications, such as matching job applicants with employers~\citep{RP99redesign}, schools with students~\citep{AS03school}, and organ donors with patients~\citep{RSU04kidney}. A fundamental solution concept in this literature is \emph{stability}~\citep{GS62college}, which requires that two unmatched agents on opposite sides should not prefer each other over their current matches. The well-known deferred-acceptance algorithm, commonly used for finding stable matchings, is known to exhibit a bias favoring one side over the other~\citep{GS62college}. This has led to a large body of work on developing fair algorithms for finding stable matchings~\citep{MW71stable,ILG87efficient,GI89stable,K97stable,TS98geometry, STQ06many,NBN22achieving}.

Somewhat surprisingly, the Nash welfare objective was not applied to the two-sided problem until the recent work of~\citet{JV24maximizing}, despite being extensively studied for one-sided preferences. This work examines the computational complexity of maximizing Nash welfare when the firms have additive valuations over the workers and are subject to capacity constraints, and the workers have cardinal valuations for the individual firms. The problem is shown to be \NPh{} even when each firm's capacity is at most~$2$; note that when each firm's capacity is $1$, a simple matching computation suffices. On the algorithmic side, the authors present a $\sqrt{\texttt{OPT}}$-approximation algorithm for the two-sided Nash welfare under positive additive valuations, where $\texttt{OPT}$ denotes the optimal objective value for the given input.

The two-sided Nash welfare, as defined in~\citep{JV24maximizing} and used in our work, involves the combined geometric mean of the workers' and firms' utilities. It is important to note that optimizing the geometric mean of only one side while ignoring the preferences of the other side can lead to suboptimal outcomes; for an example, refer to Figure 1 in \citep{JV24maximizing}. Recent work by \citet{TY24fair} focuses on heuristics for finding matchings that \emph{simultaneously} maximize the geometric mean of the two sides, whenever such matchings exist.

Some other recent works study fair division under two-sided preferences. For instance,~\citet{FMS21twosided} apply concepts from fair division, such as approximate envy-freeness and maximin share, to the many-to-many matching problem.~\citet{IKS+24fair} study the problem of assigning players to teams where both sides have preferences over each other. They focus on simultaneously achieving fairness and stability guarantees. Similarly,~\citet{BLL+23fair} study a generalization of the fair division problem where an allocation is required to be fair with respect to the preferences of the agents as well as the allocator (or the owner of the resources).

Additionally, several studies have examined matroid constraints in the two-sided matching problem, specifically in the context of stable and popular matchings~\citep{F03fixed,FK16matroid,K15stable,K17popular}. However, these studies do not consider Nash welfare.

\paragraph{Utilitarian welfare.} Finally, we also mention the results for \emph{utilitarian} welfare, which is defined as the sum of agents' utilities. In the one-sided model, a utilitarian optimal allocation can be easily computed under additive valuations by assigning each item to an agent who has the highest value for it. However, for submodular valuations, no algorithm that makes a polynomial number of value queries can provide a better approximation to utilitarian welfare than $\frac{e}{e-1}$, unless $\P{}=\NP{}$~\citep{KLM+08inapproximability}. On the algorithmic side, \citet{V08optimal} gave a randomized algorithm that matches this bound and makes a polynomial number of value queries. The algorithm can also accommodate matroid constraints. It is easy to see that the algorithmic and hardness results for utilitarian welfare in the two-sided setting mimic those in the one-sided setup.

\section{Preliminaries}
\label{sec:Preliminaries}

For any $r \in \mathbb{N}$, let $[r] \coloneqq \{1,2,\dots,r\}$. For any set $R$ and a singleton $\{j\}$, we use $R + j$ to denote $R\cup \{j\}$ and $R-j$ to denote $R\setminus \{j\}$.

\subsection*{One-Sided Model}

\paragraph{Problem instance.} An instance of the one-sided problem is defined by a tuple $\langle N,M,\V,C \rangle$, where $N$ is a set of $n$ \emph{agents}, $M$ is a set of $m$ \emph{items} (or goods), $\V = \{v_1,\dots,v_n\}$ is a set of \emph{valuation functions} and $C = (c_1,\dots,c_n)$ is a vector of \emph{capacities}. The valuation function $v_{i} : 2^{M}\to \QN$ %
specifies agent $i$'s value $v_i(S)$ for any subset $S \subseteq M$ of items. Note that we assume the valuations to be nonnegative rational numbers. The capacity $c_i \in \ZP$ is a positive integer that represents the maximum number of items that agent $i$ can receive. We will write $v_i(j)$ instead of $v_i\{j\})$ for a singleton $\{j\}$.

\paragraph{Allocation.} A partial allocation $A = (A_1,A_2,\dots,A_n)$ refers to an $n$-subpartition of the set of items $M$ such that, for all $i \neq j$, $A_i \cap A_j = \emptyset$ and $\cup_{i=1}^{n} A_{i} \subseteq M$. Here, $A_i$ represents the subset of items or the \emph{bundle} assigned to agent $i$. A partial allocation is called \emph{complete} if $\cup_{i=1}^{n} A_{i} = M$. We will use the term ``allocation'' to denote a complete allocation, and explicitly write ``partial allocation'' otherwise. We will call a partial allocation \emph{feasible} if, for each agent $i\in N$, we have $|A_i| \leq c_i$. %
Given a partial allocation $A$, we will call $v_i(A_i)$ the \emph{utility} (or value) derived by agent $i$ under $A$.

\paragraph{Nash social welfare.} The Nash Social Welfare of a partial allocation $A$ is defined as the geometric mean of the agents' utilities, i.e.,~$\NSW(A) \coloneqq \left( \prod_{i \in N} v_i (A_i) \right)^{\nicefrac{1}{n}}.$ The Nash social welfare, or $\NSW{}$ for short, is known to be \emph{scale-free}, which means that an allocation that maximizes Nash social welfare continues to do so even if agent $i$'s valuation function $v_i(\cdot)$ is scaled by a factor $\alpha_i \geq 0$.

The computational problem associated with maximizing Nash social welfare, which we call \CapacitatedOneSidedNash{}, is defined below. We will call a partial allocation \emph{Nash optimal} if it maximizes the Nash social welfare among all feasible partial allocations for a given instance.

\problembox{\CapacitatedOneSidedNash}{An instance $\mathcal{I}=\langle N,M,\V,C \rangle $ where $N$ is the set of agents, $M$ is the set of items, $\V$ is the set of valuation functions, and $C$ is a vector representing capacity of agents.}{Compute a feasible Nash optimal partial allocation $A$.}

A special case of \CapacitatedOneSidedNash{} is when there are no capacity constraints; equivalently, each agent's capacity is equal to the number of items. We call this problem \OneSidedNash{}.

\paragraph{Valuation classes.} A valuation function $v_{i} : 2^{M}\to \QN$ is said to be \emph{monotone} if, for any pair of subsets $S,T \subseteq M$ such that $S \subseteq T$, we have ${v}_i(S) \leq {v}_i(T)$, and \emph{normalized} if ${v}_i(\emptyset) = 0$. We will assume throughout that the valuations are monotone and normalized. 

Various subclasses of monotone valuations will be of interest to us. Formally, a valuation function $v_i$ is said to be:
\begin{itemize}%
    \item \emph{additive} if, for any subset $S\subseteq M$, we have $v_i(S) = \sum_{j \in S} {v}_i(j)$,
    \item \emph{submodular} if, for any subsets $S,T \subseteq M$ such that $S \subseteq T$ and any $j \notin T$, we have that $v(S \cup \{j\}) - v(S) \geq v(T \cup \{j\}) - v(T)$, and
    \item \emph{subadditive} if, for any pair of subsets $S,T \subseteq M$, we have ${v}_i(S \cup T) \leq v_i(S) + v_i(T)$.
\end{itemize}
Observe that the containment relations among these classes of valuations are $\texttt{additive} \subseteq \texttt{submodular} \subseteq \texttt{subadditive}$.

\paragraph{Query model.} Note that we allow for combinatorial valuations, which can have an exponential-sized representation in terms of the number of items. Therefore, when analyzing algorithms, it is natural to assume an oracle access to the valuations. We will focus on \emph{value} queries in our work. Given as input a bundle $S$ and an index $i$, a value query returns agent $i$'s value for the bundle $v_i(S)$. 

We will write $\poly(n,m)$ to denote a polynomial function in $n$ and $m$. We will be interested in designing algorithms that make $\poly(n,m)$ number of value queries and have $\poly(n,m)$ running time.

\paragraph{Arithmetic model.} In the running time analysis of our algorithms, we will assume throughout that all arithmetic operations (including addition, comparison, multiplication, and division) take $\mathcal{O}(1)$ time regardless of the size of the operands.

\paragraph{$\alpha$-approximation algorithm.} Given an instance $\I$, let $\ALG(\I)$ denote the allocation returned by a given algorithm $\ALG$, and let $\OPT(\I)$ denote a Nash optimal allocation for $\I$. We say that $\ALG$ is $\alpha$-approximate if, for all problem instances $\I$, we have that $\NSW(\OPT(\I)) \leq \alpha \cdot \NSW(\ALG(\I))$. Notice that $\alpha \geq 1$.

\subsection*{Two-Sided Model}

\paragraph{Problem instance.} An instance of the two-sided problem is defined by a tuple $\langle F,W,\mathcal{V},\mathcal{W},C \rangle$, where $F$ is a set of $n$ \emph{firms}, $W$ is a set of $m$ \emph{workers}, $\V = \{v_1,\dots,v_n\}$ is a set of firms' valuation functions, $\W = \{w_1,\dots,w_m\}$ is a set of workers' valuation functions, and $C = (c_1,\dots,c_n)$ is a vector of capacities. 
The valuation function $v_{i} : 2^{W}\to \QN$ %
specifies firm $i$'s value $v_i(S)$ for any subset $S \subseteq W$ of workers. The capacity $c_i \in \ZP$ is a positive integer that represents the maximum number of workers that firm $i$ can be matched with. Every worker $j \in W$ has a valuation function $w_{j}: F \to \QN$, and $w_j(i)$ represents the value that worker $j$ associates with firm $i$.

\paragraph{Many-to-one matching.} Given an instance $\I = \langle F,W,\mathcal{V},\mathcal{W},C \rangle$, %
a \textit{many-to-one matching} for $\I$, denoted by $\mu : F \times W \rightarrow \{0,1\}$, is a function that assigns each worker-firm pair a weight of either 0 or 1 such that for every worker $j \in W$, $\sum_{i \in F} \mu(i, j) \leq 1$, and for every firm $i \in F$, $\sum_{j \in W} \mu(i, j) \leq c_i$. Further, we define $\mu_{i} := \{j \in W : \mu(i, j) = 1\}$ and $\mu_{j} := \{i \in F : \mu(i, j) = 1\}$ as the set of workers and firms matched with firm $i$ and worker $j$, respectively.

\paragraph{Nash social welfare.} The Nash Social Welfare of a many-to-one matching $\mu$ is defined as the geometric mean of the utilities of the firms and workers under $\mu$, i.e.,
\begin{center}
    $\NSW(\mu) = \left(\prod_{i \in F} v_{i}(\mu_{i}) \prod_{j \in W} w_j(\mu_j)\right)^{\frac{1}{m+n}}.$
\end{center}
Note that similar to the one-sided model, the two-sided Nash social welfare is also \emph{scale-free}.

The computational problem associated with maximizing the two-sided Nash Social Welfare, which we call \CapacitatedTwoSidedNash{}, is defined below. We will call a many-to-one matching \emph{Nash optimal} if it maximizes the Nash social welfare among all feasible many-to-one matchings for a given instance.

\problembox{\CapacitatedTwoSidedNash{}}{An instance $\mathcal{I}=\langle F,W,\V,\W,C \rangle $ where $F$ is the set of firms, $W$ is the set of workers, $\V$ is the set of firms' valuation functions, $\W$ is the set of workers' valuation functions and $C$ is a vector representing capacity of firms.}{Compute a feasible Nash optimal many-to-one matching~$\mu$.}

A special case of \CapacitatedTwoSidedNash{} is when there are no capacity constraints on the firms; equivalently, each firm's capacity is equal to the number of workers. We call this problem \TwoSidedNash{}. The concepts of an $\alpha$-approximation algorithm, query model, and valuation classes for the two-sided model are defined analogously to the one-sided setting. In the two-sided case, value queries can be made for both workers and firms.

\paragraph{The case of zero Nash welfare and inadequate capacities.}
In the \CapacitatedTwoSidedNash{} problem, we will assume that the capacities are adequate, that is, $\sum c_i \geq m$.
Note that there can be situations, in both the \CapacitatedOneSidedNash{} and \CapacitatedTwoSidedNash{} problems, where the optimal solution has zero Nash welfare. In this case however, an arbitrary feasible solution satisfies the required approximation ratio. Therefore, we will assume in the rest of the paper that the optimal solution has a nonzero Nash welfare.

\section{Approximation Algorithm for the One-Sided Model}
\label{sec:OneSidedAlgo}
In this section, we will show that given any $\eps > 0$, there is a $(6+\eps)$- approximation algorithm for \CapacitatedOneSidedNash{} under submodular valuations. The running time of the algorithm is strongly polynomial in $n$ and $m$, i.e., it doesn't depend on the size of the valuations. Our algorithm and its analysis are obtained by modifying the algorithm of \citet{GHL+23approximating} that gives a $(4+\eps)$-approximation under submodular valuations for the \OneSidedNash{} problem, i.e., the one-sided Nash welfare maximization problem \emph{without} capacity constraints.

Before discussing our algorithm, it is relevant to discuss some natural approaches that do not work.

\paragraph{Limitations of natural approaches.} The algorithm of~\citet{GHL+23approximating} is not designed to handle capacities and may return infeasible allocations. Additionally, it is possible to design a family of instances in which the optimal Nash welfare under capacities is arbitrarily smaller than the optimal Nash welfare without capacities\footnote{A simple example is given by an instance with $n$ agents and $kn$ items where each agent values each item at $c$ and all valuations are additive. The uncapacitated maximum $\NSW$ is $kc$. If the capacity of each agent is just $1$, then the capacitated maximum $\NSW$ is $c$.}. In such instances, a $(4+\eps)$-approximation algorithm for the unconstrained problem would decidedly return an infeasible allocation.

Another common approach for handling constraints is by \emph{modifying} the valuation functions. For example, an instance with \emph{additive} valuations and capacity constraints can be converted to an unconstrained instance with \emph{submodular} valuations via the following transformation: For any set $S$ of items, define agent $i$'s value $v_i(S)$ as the sum of values of the $\min\{c_i,|S|\}$ most valuable items in $S$, where $c_i$ is the capacity for agent $i$. The resulting valuation function $v_i(\cdot)$ can be observed to be submodular. Consequently, the algorithm of \citet{GHL+23approximating}, which gives a $(4+\eps)$ approximation in the unconstrained instance, gives a similar approximation for the original additive valuations instance with capacity constraints. Unfortunately, when the given instance has \emph{submodular} valuations with capacity constraints, such a transformation may not result in an unconstrained submodular instance, as demonstrated by the following example.
\begin{example}
Let $v$ be a submodular valuation function of an agent with capacity $c$. Let us define a new valuation function $v'$ such that $v'(S) \coloneqq \max_{S'\subseteq S : |S'| \leq c} v(S')$. Then, $v'$ can fail to be submodular. Consider a set of four items $M = \{w_1, w_2, w_3, w_4\}$ and say $c = 2$. Suppose the agent's values under $v(\cdot)$ are:
\begin{itemize}%
    \item $v(\emptyset) = 0$,
    \item $v(\{w_2,w_4\}) = 4$ and $v(\{w_1, w_3,w_4\}) = 3$, and
    
    \item for all other subsets $S \subseteq M$, $v(S) = \min\{4, |S|+1\}$.
\end{itemize}
It can be observed that $v(\cdot)$ is submodular and monotone.

Note that $v'(\{w_1,w_3\})  = 3$ and $v'(\{w_1,w_3,w_4\}) = 3$, implying a marginal increase of $0$ upon adding $w_4$. Similarly, $v'(\{w_1,w_2, w_3\}) = 3$ and $v'(\{w_1,w_2,w_3, w_4\}) = 4$, which gives a marginal increase of $1$ upon adding $w_4$. Since adding $w_4$ to a smaller set results in a smaller marginal, we get that $v'(\cdot)$ is not submodular.\hfill{}\qed
\end{example}

Having motivated the challenge of approximating Nash welfare in the capacitated setting, let us now formally state our main result.

\begin{restatable}[]{theorem}{OneSidedApprox}
For any $\eps > 0$, there exists a $(6+\eps)$-approximation algorithm for \CapacitatedOneSidedNash{} under submodular valuations that runs in $\poly(n,m)$ time and makes a $\poly(n,m)$ number of value queries.
\label{theorem:capacitated-1-sided}
\end{restatable}

\begin{remark}[Approximate fairness]
    In the unconstrained setting under additive valuations, it is known that any Nash optimal allocation, say $A$, is \emph{envy-free up to one item} (EF1)~\citep{CKM+19unreasonable}. This property entails that for any pair of agents $i,k$, there exists some item $g \in A_j$ such that $v_i(A_i) \geq v_i(A_k \setminus \{g\})$. However, this implication breaks down under capacity constraints. Recently,~\citet{WLG21budget} studied \emph{budget constraints} in the one-sided problem, of which capacity constraints are a special case. They showed that under subadditive valuations, any budget-feasible $\alpha$-approximate allocation for Nash welfare satisfies $\frac{1}{4\alpha}$-EF1. In the presence of capacity constraints, this property requires that for any pair of agents $i,k$ and any subset $S \subseteq A_k$ such that $|S| \leq c_i$, there exists some item $g \in S$ such that $v_i(A_i) \geq \frac{1}{4\alpha} v_i(S \setminus \{g\})$. By combining their result with the guarantee in \Cref{theorem:capacitated-1-sided}, we obtain that the allocation returned by our algorithm additionally satisfies $\frac{1}{(24+4\eps)}$-EF1.\qed
\end{remark}

Towards proving \Cref{theorem:capacitated-1-sided}, we will solve a related problem, \ExactCapacitatedOneSidedNash{}, in which each agent is assigned \emph{exactly} as many items as its capacity. Formally, an allocation $A = (A_i)_{i\in N}$ is feasible for a given instance of \ExactCapacitatedOneSidedNash{} if, for every $i \in N$, we have $|A_i| = c_i$. It is easy to construct an approximation-preserving reduction between the two problems by adding dummy items that do not affect the value of any subset of the original items. Thus, \Cref{theorem:capacitated-1-sided} follows from \Cref{lemma:exact-capacitated-1-sided} below.

\begin{lemma}
For any $\eps > 0$, there exists a $(6+\eps)$ approximation algorithm for \ExactCapacitatedOneSidedNash{} under submodular valuations that runs in $\poly(n,m)$ time and makes $\poly(n,m)$ number of value queries.
\label{lemma:exact-capacitated-1-sided}
\end{lemma}

The algorithm in \Cref{lemma:exact-capacitated-1-sided} is a modification of that of \citet{GHL+23approximating} and works in three phases. The first phase involves computing a \emph{one-to-one maximum weight matching} $\tau$ of the agents and items. We denote the set of items matched in this phase by $H$. Let $J \coloneqq M \setminus H$ be the set of remaining items. In the second phase, the items in $J$ are allocated through \emph{local search}. Crucially, we only allow for \emph{two-way transfers} (or swaps) between pairs of agents to ensure that the capacity constraints are not violated during the course of the local search. In the final phase, the items in $H$ are \emph{rematched} to the agents. Note that the first and the third phases of our algorithm are identical to that of \citet{GHL+23approximating}. In the second phase, the algorithm of \citet{GHL+23approximating} performs local search via \emph{one-way} transfer of items, whereas our algorithm uses \emph{two-way} exchanges to maintain the capacity constraints.

In the following, we will define some terminology that will be used in describing our algorithm for \ExactCapacitatedOneSidedNash{}. %

\paragraph{Endowed valuation functions.} In the local search phase of the algorithm, we do not use the actual valuation functions of the agents; instead, we use the \emph{endowed} valuation functions. Let $\overline{N} \coloneqq \{ i \in N: v_i(J) > 0 \}$ be the set of agents that assign a positive value to the set $J$. For every agent $i\in \overline{N}$, we define the endowed valuation function $\overline{v}_i$ as $\overline{v}_i(S) \coloneqq v_i(S) + v_i(\ell(i))$ for any subset $S\subseteq M$, where $\ell(i)$ is the favourite item of agent $i$ in $J$. Note that if $v_i$ is submodular, then $\overline{v}_i$ is also submodular.

\paragraph{Accuracy parameter.} Our local search subroutine will use the parameter $\overline{\eps} = -1 + (1+\eps)^{1/m}$ such that the minimum multiplicative increase in $\NSW$ required to perform a swap is $1+\overline{\eps}$.

\paragraph{Swaps in local search.} Our local search phase uses swaps (or item exchanges) between agents whenever the swap provides a large enough increase in $\NSW$. Note that two-way swaps maintain capacity constraints. Specifically, we will consider the following two kinds of swaps.
\begin{itemize}%
    \item \ShufflingSwap{}: Let $\cR = (R_i)_{i\in \overline{N}}$ be a partial allocation of the items in $J$ among the agents in $\overline{N}$. A $\ShufflingSwap$ is said to exist if there is an item $j\in R_i$ allocated to an agent $i\in \overline{N}$ and an item $k\in R_{i'}$ allocated to an agent $i' \in \overline{N}$ such that $\left(\frac{\overline{v}_{i}(R_{i}-j+k)  \overline{v}_{i'}(R_{i'}-k+j)}{\overline{v}_i(R_{i})\overline{v}_{i'}(R_{i'})}\right)^{\nicefrac{1}{n}} > 1+\overline{\eps}$. The algorithm performs this swap by allocating $j$ to $i'$ and $k$ to $i$.
    \item \AlteringSwap{}: Let $\cR = (R_i)_{i\in \overline{N}}$ be a partial allocation of the items in $J$ among the agents in $\overline{N}$. A $\AlteringSwap$ is said to exist if there is an unallocated item $k\in J$ and an item $j\in R_i$ allocated to an agent $i\in \overline{N}$ such that $\left(\frac{\overline{v}_{i}(R_{i}-j+k)}{\overline{v}_{i}(R_{i})}\right)^{\nicefrac{1}{n}} > 1+\overline{\eps}$. We perform this swap by allocating $k$ to $i$ and $j$ is left unallocated.
\end{itemize}

\paragraph{$\overline{\eps}$-local optimum.} A partial allocation $\cR = (R_i)_{i\in \overline{N}}$ of $J$ to $\overline{N}$ is an \emph{$\overline{\eps}$-local optimum} with respect to the endowed valuations $\overline{v}_i$ if there is no \ShufflingSwap{} or \AlteringSwap{} possible. 

With the terminology in place, we formally describe our algorithm in \Cref{alg:NSW-template} and the local search subroutine in \Cref{alg:localSearch}. It should be noted that when the Nash optimal partial allocation has a Nash welfare of $0$, the output of our algorithm trivially satisfies the desired approximation guarantee. Hence, we will assume in our analysis that the optimal allocation has a nonzero Nash welfare.

\begin{algorithm}[t]
\DontPrintSemicolon
\KwIn{An instance $\langle N,M,\V,C \rangle$ of \ExactCapacitatedOneSidedNash{}.}
\KwOut{A partial allocation $A=(A_i)_{i\in N}$.}

\tcc{Phase 1: Find a maximum weight one-to-one matching}
Compute a one-to-one matching $\tau: N \to M$ that maximizes $\prod_{i\in N} v_i(\tau(i))$.

\tcc{$H$ and $J$ are the allocated and unallocated items respectively.} 
Set $ H \coloneqq \{ \tau(i) : i \in N \} $ and $J \coloneqq M \setminus H$. 

\tcc{Phase 2: Local search using the unallocated items}
Compute the allocation $\mathcal{R} \coloneqq $\pLocalSearch{$\langle N, J, \V, C \rangle$}.\;

\tcc{Phase 3: Rematching the previously allocated items in $H$}
Find a matching $\delta: N \to  H$ maximizing $\prod_{i \in N} v_i(R_i + \delta(i))$. \label{algo:phase3}\; 
\Return{$A =(R_i + \delta(i))_{i\in N}$}\;
\caption{Approximating \ExactCapacitatedOneSidedNash{} for submodular valuations}
\label{alg:NSW-template}
\end{algorithm}

\begin{algorithm}[ht]
\DontPrintSemicolon
\KwIn{An instance $\langle N, J,\V, C\rangle$.}
\KwOut{A partial allocation $\cR \coloneqq (R_i)_{i\in N}$.}

\tcc{The set of agents which have positive utility for $J$.}
$\overline{N} \gets \{ i \in    N: v_i(J) > 0 \}$ \;

\tcc{$\ell(i)$ is the favourite item of $i$ in $J$}
$\ell(i) \gets \argmax \{ v_i(\ell): \ell \in J \}$ for $i \in \overline{N}$\;

Define the endowed valuations by $\overline{v}_i(S) \coloneqq v_i(\ell(i)) + v_i(S)$ for all $i \in \overline{N}, S\subseteq M$.\;

Pick an arbitrary partial allocation $\cR$ of items in $J$ to agents in $\overline{N}$ such that $|R_{i}| = c_i-1$ for every agent $i \in \overline{N}$. Keep all other items in $J$ unallocated.\;

\While{there exists a \ShufflingSwap{} or \AlteringSwap{} with respect to $\overline{v}(\cdot)$}{
perform the corresponding swap}

Extend the partial allocation $\cR$ to all the agents in $N$ by arbitrarily allocating the unallocated items to agents in $N \setminus \overline{N}$ such that $|R_i| = c_i -1$ for every agent $i \in N \setminus \overline{N}$.\;

\label{alg:line}

\Return{$\cR \coloneqq (R_i)_{i\in N}$}
\caption{{\texttt{LocalSearch}}}\label{alg:localSearch}
\end{algorithm}

Our next lemma shows that the local search subroutine (corresponding to Phase 2 in \Cref{alg:NSW-template}) converges to a feasible $\overline{\eps}$-local optimum in polynomial time.

\begin{restatable}[Local search converges efficiently]{lemma}{LocalSearchConvergence}
    For any given instance $\langle N, J, \V, C \rangle$, \Cref{alg:localSearch} terminates in $\mathcal{O} \left( \frac{m}{\eps}\log m \right)$ iterations and returns a partial allocation $(R_i)_{i\in N}$ that satisfies $|R_i| = c_i -1$ for all $i \in N$. Moreover, the partial allocation $(R_i)_{i \in \overline{N}}$ computed just before \Cref{alg:line} is an $\overline{\eps}$-local optimum.
     \label{lemma:local}
\end{restatable}

\begin{proof} The local search subroutine terminates only when no more swaps are possible. Thus, the partial allocation $\cR$ returned by the local search subroutine is an $\overline{\eps}$-local optimum. Note that the initial partial allocation satisfies $|R_i| = c_i -1$ for all $i\in \overline{N}$ and any subsequent swaps do not change the size of $R_i$. Also, we allocate $c_i-1$ items at the end for all agents $i \in N \setminus \overline{N}$. Hence, the final partial allocation satisfies $|R_i| = c_i -1 $ for all $i \in N$. 

    Now note that $v_i(J) \leq (|J| +1)\overline{v}_i(\emptyset) \leq m \overline{v}_i(\emptyset)$ using submodularity and the definition of endowed valuations. Hence,
    $$ \left(\prod_{i \in \overline{N}} \overline{v}_i(J)\right)^{\frac{1}{n}} \leq m\left(\prod_{i \in \overline{N}} \overline{v}_i(\emptyset)\right)^{\frac{1}{n}},$$
    which shows that the product $\left(\prod_{i \in \overline{N}} \overline{v}_i(R_i)\right)^{\frac{1}{n}}$ can only increase by at most a factor of $m$. Each swap increases this product by a factor of $(1 + \overline{\eps})$. Hence, the total number of iterations is bounded by $\log_{(1+\overline{\eps})}m = \mathcal{O}(\frac{m}{\eps}\log m)$.
\end{proof}

In a manner similar to~\citet{GHL+23approximating}, we will define a \emph{price} $p_{jk}$ for every pair of items $j, k$ after \Cref{alg:localSearch} has returned a partial allocation $\cR$ of items in $J$ among agents in $N$. Let $j, k \in J$. Then,
\[p_{jk} \coloneqq 
    \begin{cases} 
      \frac{\max \{ 0, 
 \overline{v}_i(R_{i})-\overline{v}_i(R_i-j+k) \}}{\overline{v}_i(R_i-j+k)} & \text{ if } j \in R_i,k \in R_{i'} \text{ and } i, i' \in \overline{N}  \\
      0 & \text{otherwise.} 
    \end{cases}\]

Next, we will present some bounds on the prices.

\begin{restatable}[Price bound]{lemma}{PriceBound}
Let $i \in N$ be an agent and for all $j \in R_i$, let $k_j$ be any item in $J$. Then, $\sum_{j \in R_i}p_{jk_j} \le 1$. In particular, this implies that $p_{jk} \leq 1$ for all $j \in R_i, k \in J$.
\label{lemma:price-sum}
\end{restatable}

\begin{proof}
If $i \in N\setminus \overline{N}$, $p_{jk} = 0$ for all $j \in R_i, k \in J$, and the statement trivially holds true. Hence, we assume $i \in \overline{N}$. Using monotonicity, we can get the following bound:
\begin{align*}
\max(0, \overline{v}_i(R_i) - \overline{v}_i(R_i-j+k))  &= \max(0, v_i(R_i) - v_i(R_i-j+k))\\
& \le v_i(R_i) - v_i(R_i - j).
\end{align*}

Additionally,
\begin{align*}
\overline{v}_i(R_i - j + k) &= v_i(R_i-j+k)+v_i(\ell(i))\\ 
& \ge v_i(R_i - j) + v_i(j) & \textnormal{(by monotonicity)}\\
&\ge v_i(R_i) & \textnormal{(by submodularity)}.
\end{align*}

Therefore, we have:
\[
        \sum_{j \in R_i} p_{j k_j} \leq \sum_{j \in R_i} \dfrac{v_i(R_i) - v_i(R_i - j)}{v_i(R_i)} \leq 1,
\]
where we have used the inequality $v_i(R_i) \geq \sum_{j \in R_i} v_i(R_i) - v_i(R_i - j)$, which follows from submodularity.
\end{proof}

For analysis of the local search, we define $\hat{\eps} = (1+\overline{\eps})^n-1$ where $\overline{\eps}$ is the accuracy parameter of the local search. Now, we write the termination condition of the local search in terms of prices.

\begin{restatable}[Termination condition]{lemma}{LocalSearchTermination}
Let item $k \in R_i$ for some agent $i \in \overline{N}$ and let item $j \in J$. Then,
\begin{center}
    $\frac{\overline{v}_i(R_i-k+j)}{\overline{v}_i(R_i)} \leq (1+\hat\eps)(1+p_{jk})$.
\end{center}
\label{lemma:termination}
\end{restatable}

\begin{proof}
If $j \in R_i$, the statement holds true since the left-hand side is at most $1$, and the right-hand side equals $1 + \hat\eps$. Hence, we assume $j \notin R_i$ and consider two cases, each relating to the two kinds of swaps.

\textbf{Case 1: } $j$ does not belong to an agent in $\overline{N}$. 

The fact that the local search terminates implies that a \AlteringSwap{} is not possible between $k$ and $j$. Hence,

$$\frac{\overline{v}_i(R_i-k+j)}{\overline{v}_i(R_i)} \leq (1+\hat\eps) \leq (1+\hat\eps)(1+p_{jk})$$ since $p_{jk}$ is nonnegative.

\textbf{Case 2: } $j$ belongs to an agent $i' \in \overline{N}$,  such that $i' \neq i$.

Now, the local search termination implies that a \ShufflingSwap{} is not possible between $k$ and $j$. Hence,

$$\frac{\overline{v}_i(R_{i}-k+j)\overline{v}_{i'}(R_{i'}-j+k)}{\overline{v}_i(R_{i})\overline{v}_{i'}(R_{i'})} \leq (1+\hat\eps)$$

Since endowed valuations are always positive for an agent in $\overline{N}$, we can rearrange some terms to get

$$\frac{\overline{v}_{i}(R_{i}-k+j)}{\overline{v}_{i}(R_{i})} \leq (1+\hat\eps)\left(\frac{\overline{v}_{i'}(R_{i'})}{\overline{v}_{i'}(R_{i'}-j+k)}\right) \leq (1+\hat\eps)(1 + p_{jk}).$$
\end{proof}

Next, we consider an arbitrary set $T \subseteq J$ of size $|R_i|+1$ and show that the increase in value provided by such a set can be bounded using prices.

\begin{restatable}[Bounded increase in value]{lemma}{BoundedIncreaseInValue}
Let $i\in \overline{N}$ be any agent and let $T \subseteq J$ such that $|T| \leq |R_{i}|+1$. Let $k_j \in R_i$ be a unique item for each $j \in T$, possibly with one repetition. Then,
\begin{center}
    $ \frac{\overline{v}_i(R_i \cup T)}{\overline{v}_i(R_i)} \leq 2 + \sum_{j \in T}(2\hat \eps+p_{jk_{j}})$.
\end{center}
\label{lemma:sum_bound}
\end{restatable}

In proving \Cref{lemma:sum_bound}, we will use a fact about submodular endowed valuation functions from~\citep{GHL+23approximating}.

\begin{restatable}[\citet{GHL+23approximating}]{lemma}{marginals}
Let $\overline{v}: 2^{M} \to \ZN$ be a submodular endowed valuation. Let $R\subseteq J$ be any set. Then, for any $j \in R$,
\begin{center}
    $\overline{v}(R-j) \ge  \sum_{k \in R} (\overline{v}(R) - \overline{v}(R-k))$.
\end{center}
\label{lemma:marginals}
\end{restatable}

\begin{proof} (of \Cref{lemma:sum_bound}) Using submodularity, we can write
\begin{align*}
\frac{\overline{v}_i(R_i \cup T)}{\overline{v}_i(R_i)} & \leq \frac{\overline{v}_i(R_i)+\sum_{j \in T}(\overline{v}_i(R_i+j)-\overline{v}_i(R_i))}{\overline{v}_i(R_i)}\\
& \leq \frac{\overline{v}_i(R_i)+\sum_{j \in T}(\overline{v}_i(R_i+j-k_j)-\overline{v}_i(R_i-k_j))}{\overline{v}_i(R_i)}.
\end{align*}

We will now use the condition of the local search (\Cref{lemma:termination}) combined with the previous inequality to get
\begin{equation*}
\begin{split}
    \frac{\overline{v}_i(R_i \cup T)}{\overline{v}_i(R_i)} &\leq  1+ \sum_{j \in T}\left((1+\hat \eps)(1+p_{jk_j})-\frac{\overline{v}_i(R_i-k_j)}{\overline{v}_i(R_i)}\right) \\ & =    1+ \sum_{j \in T}\left((1+\hat \eps)(1+p_{jk_j})+\frac{\overline{v}_i(R_i) - \overline{v}_i(R_i-k_j)}{\overline{v}_i(R_i)}  -1 \right).
\end{split}  
\end{equation*}
 
If $R_i$ is empty, then note that $\frac{\overline{v}_i(R_i) - \overline{v}_i(R_i-k_j)}{\overline{v}_i(R_i)} = 0$. Therefore, expanding $(1+\hat \eps)(1+p_{jk_j})$ and using the fact that $p_{jk_j} \leq 1$ as shown in \Cref{lemma:price-sum} we obtain the desired inequality as follows.

\begin{equation*}
\begin{split}
    \frac{\overline{v}_i(R_i \cup T)}{\overline{v}_i(R_i)} &\leq  1+ \sum_{j \in T}\left((1+\hat \eps)(1+p_{jk_j})-1\right) \\ & \leq    2+ \sum_{j \in T}(2\hat \eps +p_{jk_j} ). 
\end{split}  
\end{equation*}

Otherwise, if $R_{i}$ is not empty, note that the size of $T$ can be atmost 1 more than the size of $R_{i}$ and that we have a unique $k_j$ in $R_{i}$ for all but one $j$. Let that one repeated item be $\overline{k} \in R_{i}$. Hence, the previous inequality can be written as 
\begin{equation*}
    \begin{split}
        \frac{\overline{v}_i(R_i \cup T)}{\overline{v}_i(R_i)} \leq  1 & + \sum_{j \in T}((1+\hat \eps)(1+p_{jk_j})-1) \\ &+ \sum_{k \in R_{i}}\frac{\overline{v}_i(R_i)-\overline{v}_i(R_i-k)}{\overline{v}_i(R_i)} + \frac{\overline{v}_i(R_i)-\overline{v}_i(R_i-\overline{k})}{\overline{v}_i(R_i)}
    \end{split}
\end{equation*}

We can use \Cref{lemma:marginals} to bound the second summation as follows:

$$\sum_{k \in R_{i}}\frac{\overline{v}_i(R_i)-\overline{v}_i(R_i-k)}{\overline{v}_i(R_i)} \leq \frac{\overline{v}_i(R_i - \overline{k})}{\overline{v}_i(R_i)}$$

We can combine this bound with the last term in the previous inequality to get
\begin{equation*}
    \begin{split}
        \frac{\overline{v}_i(R_i \cup T)}{\overline{v}_i(R_i)} &\leq  1+ \sum_{j \in T}((1+\hat \eps)(1+p_{jk_j})-1)+\frac{\overline{v}_i(R_i)}{\overline{v}_i(R_i)} \\
        & = 2 + \sum_{j \in T}((1+\hat \eps)(1+p_{jk_j})-1)  \leq 2 + \sum_{j \in T}(2\hat \eps+p_{jk_j}).\qedhere
    \end{split}
\end{equation*}

where we use \cref{lemma:price-sum} in the last inequality.

\end{proof}

 We will now derive a bound similar to \Cref{lemma:sum_bound} for \emph{actual} valuations rather than endowed valuations.

\begin{restatable}[Bounded increase in actual value]{lemma}{BoundedIncreaseInActualValue}
Let $i\in \overline{N}$ and let $S \subseteq J$ be any set of items such that $|S| \leq |R_i| +1$. Let $k_j \in R_i$ be a unique item for each $j \in S$, possibly with one repetition. Then,
\begin{center}
    $\frac{v_i(S)}{\max \{v_i(R_i), v_i(\ell(i)\}} \leq 3+  2 \sum_{j \in S} (2\hat{\eps} + p_{jk_j})$.
\end{center}
\label{lemma:price-bound}
\end{restatable}

\begin{proof}
By ~\Cref{lemma:sum_bound},
$$ \frac{v_i(\ell(i)) + v_i(S)}{v_i(\ell(i)) + v_i(R_i)} = \frac{\overline{v}_i(S)}{\bar v_i(R_i)} 
\leq \frac{\bar v_i(R_i \cup S)}{\bar v_i(R_i)} \leq 2 + \sum_{j \in S} (2\hat{\eps} + p_{jk_j}). $$

Let $\lambda = \frac{v_i(R_i)}{v_i(\ell(i))}$. We can rewrite the inequality above as follows:
$$  \frac{1 + \frac{v_i(S)}{v_i(\ell_i)}}{1 + \lambda} \leq 2 + \sum_{j \in S} (2\hat{\eps} + p_{jk_j}). $$

If $v_i({\ell_i}) \geq v_i(R_{i})$, or equivalently, $ 0 \leq \lambda \leq 1$, we have the following inequality:

\begin{align*}
\frac{v_i(S)}{\max\{ v_i(\ell(i)), v_i(R_i)\}} & = \frac{v_i(S)}{v_i(\ell_i)}  \leq (1+\lambda) (2 + \sum_{j \in S} (2\hat{\eps} + p_{jk_j})) - 1 \\
& \leq 2\lambda +1+ (1+\lambda) \sum_{j \in S} (2\hat{\eps} + p_{jk_j}).
\end{align*}

If $v_i({\ell_i}) < v_i(R_{i})$, or equivalently $\lambda > 1$, we divide by $\lambda$ to obtain:

$$\frac{v_i(S)}{\max\{ v_i(\ell(i)), v_i(R_i)\}} = \frac{v_i(S)}{v_i(R_i)} \leq \frac{1}{\lambda} + 2+ ({1}/{\lambda} + 1) \sum_{j \in S} (2\hat{\eps} + p_{jk_j}).$$

In both inequalities, the maxima is attained at $\lambda = 1$, which gives
$$\frac{v_i(S)}{\max\{ v_i(\ell(i)), v_i(R_i)\}} \leq 3 + 2 \sum_{j \in S} (2\hat{\eps} + p_{jk_j}),$$
as desired.
\end{proof}

Given a matching $\rho: N \to H \cup \{\emptyset\}$, we will define $(\cR,\rho)$ as a partial allocation in which every agent receives the set $R_i \cup \rho(i)$. The $\NSW$ of this allocation will be:
$$\NSW(\mathcal{R},\rho)\coloneqq \prod_{i\in N}v_i(R_i +  \rho(i))^\frac{1}{n}.$$

Recall that, in Phase 3, our algorithm computes a matching $\delta$ such that $\NSW(\cR,\delta)$ is maximized. If we prove that there exists a matching $\sigma: N \to H \cup \{\emptyset\}$ such that $\NSW(\cR,\sigma) \geq \frac{\NSW(\OPT)}{6(1+\eps)}$, then we will be done. To prove the existence of such a matching $\sigma$, we first define a mapping $\cT: N \rightarrow 2^M$, which may not correspond to a valid partial allocation. However, this mapping is defined only for the purpose of the analysis, and since each agent is mapped under $\cT$ to a bundle, we can nevertheless define $\NSW$ of $\cT$. We will show that $\NSW(\cT) \geq \frac{\NSW(\OPT)}{6(1+\eps)}$, and then prove that a matching $\sigma$ satisfying $\NSW(\cR, \sigma) \geq \NSW(\cT) \geq \frac{\NSW(\OPT)}{6(1+\eps)}$ exists.

Consider an optimal partial allocation $\OPT$ of the given \CapacitatedOneSidedNash{} instance. Further, let $S_i \coloneqq \OPT_i \cap \, J$ and $H_i \coloneqq \OPT_i \cap \, H$. For every agent $i\in N$, define $g(i)$ to be the item in $H_i$ which provides the maximum marginal gain when added to $S_i$. Formally, $ g(i) \coloneqq \argmax_{j\in H_i} v_i(S_i + j)$. If $H_i$ is empty, we define $g(i) \coloneqq \emptyset$.

Consider the following partitioning of the set of agents $N$:
\begin{itemize}%
    \item $N_{g} \coloneqq \left\{ i \in N: v_i(g(i)) \ge \max\left\{ v_i(R_i), v_i(\ell(i))\right\}\right\}$
    \item $N_{R} \coloneqq \left\{ i \in N \setminus N_{g}: v_i(R_i) \ge \max\left\{ v_i(g(i)), v_i(\ell(i))\right\}\right\}$
    \item $N_{\ell} \coloneqq N \setminus (N_{g}\cup N_R)$.
\end{itemize}

The intermediate mapping $\cT = (T_i)_{i\in N}$ is defined as follows:
\[
T_i:=\begin{cases}
\{g(i)\}\, ,&\mbox{if }i\in N_{g}\, ,\\
R_i \, ,&\mbox{if }i\in N_{R}\, ,\\
\{\ell(i)\}\, ,&\mbox{if }i\in  N_{\ell}\, .\\
\end{cases}
\]
Note that the mapping $\cT$ may not induce a feasible partial allocation because the item $\ell(i)$ may not be unique for each agent and may even be contained in some $R_{i'}$.

\begin{restatable}[]{lemma}{BoundForInfeasibleAllocation}
    The mapping $\cT$ satisfies $\NSW(\cT) \geq \frac{\NSW(\OPT)}{6(1+\eps)}$.
    \label{lemma:intermediate}
\end{restatable}

\begin{proof}
We can use our partition of agents to write 
\[
\frac{\NSW{\OPT}}{\NSW(\cT)} =
\prod_{i \in N_{g}} \left(\frac{v_i(\OPT_i)}{v_i(g(i))}\right)^{\frac{1}{n}}
 \prod_{i \in N_R} \left(\frac{v_i(\OPT_i)}{v_i(R_i)}\right)^{\frac{1}{n}} 
\prod_{i \in N_{\ell}} \left(\frac{v_i(\OPT_i)}{v_i(\ell(i))}\right) ^{\frac{1}{n}}.\, \]

Consider any agent $i \in N \setminus \overline{N}$. Observe that this agent must be in $N_g$. Hence, for such an agent,
\begin{equation*}
\begin{split}
    \frac{v_i(\OPT_i)}{v_i(g(i))} = \frac{v_i(S_i \cup H_i)}{v_i(g(i))} \leq  \frac{|H_i|v_i(g(i))}{v_i(g(i))} =  |H_i|.
\end{split}
\end{equation*}

Now consider any agent $i \in \overline{N}$. This agent may belong to either $N_g,N_R$ or $N_{\ell}$. We handle each case separately.

If $i \in N_g$,
\begin{align*}
    \frac{v_i(\OPT_i)}{v_i(g_i)}
    & \le \frac{v_i(S_i \cup H_i)}{v_i(g(i))}
    \le \frac{v_i(S_i) + |H_i|v_i(g(i))}{v_i(g(i))}\\
    & = \frac{v_i(S_i)}{v_i(g(i))} + |H_i|
    \le \frac{v_i(S_i)}{\max\{v_i(\ell(i)), v_i(R_i)\}} + |H_i|.
\end{align*}

If $i \in N_R$, we get
\begin{align*}
    \frac{v_i(\OPT_i)}{v_i(R_i)}
    & \le \frac{v_i(S_i \cup H_i)}{v_i(R_i)}
    \le \frac{v_i(S_i) + |H_i|v_i(g(i))}{v_i(R_i)}\\
    & \le \frac{v_i(S_i)}{v_i(R_i)} + |H_i|
    \le \frac{v_i(S_i)}{\max\{v_i(\ell(i)), v_i(R_i)\}} + |H_i|.
\end{align*}

Similarly, if $i \in N_{\ell}$, we get
\begin{align*}
    \frac{v_i(\OPT_i)}{v_i(\ell_i)}
    & \le \frac{v_i(S_i \cup H_i)}{v_i(\ell_i)}
    \le \frac{v_i(S_i) + |H_i|v_i(g(i))}{v_i(\ell_i)}\\
    & \le \frac{v_i(S_i)}{v_i(\ell_i)} + |H_i|
    \le \frac{v_i(S_i)}{\max\{v_i(\ell(i)), v_i(R_i)\}} + |H_i|.
\end{align*}

Hence, for any agent $i \in \overline{N}$,

$$\frac{v_i(\OPT_i)}{v_i(T_i)} \leq \frac{v_i(S_i)}{\max\{v_i(\ell(i)), v_i(R_i)\}} + |H_i|.$$

Therefore,
\begin{equation*}
\prod_{i \in N} \left(\frac{v_i(T_i)}{v_i(\sigma_i)}\right) \leq \prod_{i \in N\setminus \overline{N}} |H_i|
\prod_{i \in \overline{N}} \left(\frac{v_i(S_i)}{\text{max}(v_{i}(l_{i}),v_i(R_{i}))} + |H_{i}|\right)\,
\end{equation*}

Note that $|S_i| \leq c_i$ because of the capacity constraints on $\OPT$. Also, $|R_i| = c_i-1$ because the swaps performed during the local search preserve the cardinality of the initial partial allocation. Hence, we can use \Cref{lemma:price-bound} in this inequality.

$$\prod_{i \in N} \left(\frac{v_i(\OPT_i)}{v_i(T_i)}\right) \leq \prod_{i \in N\setminus \overline{N}} |H_{i}| \prod_{i \in \overline{N}} (3 + 2 \sum_{j \in S_i}
(2\eps+p_{jk_j})+|H_i|),$$

where we choose $k_j$ such that for each $S$, $k_j$ is a unique item for each $j$, possibly with one repetition.

Now applying AM-GM inequality gives us
\begin{align*}
  \prod_{i \in N} \left(\frac{v_i(\OPT_i)}{v_i(T_i)}\right) \leq \frac{1}{n^{n}} \left(\sum_{i \in N\setminus  \overline{N}}|H_{i}| + \sum_{i \in \overline{N}}(3+2\sum_{j \in S_{i}}(2\eps+p_{jk_j})+|H_{i}|)\right)^{n}.
\end{align*}

First we bound the sum of prices:

$$\sum_{i \in \overline{N}} \sum_{j \in S_{i}}p_{jk_j} \leq  \sum_{j \in J}p_{jk_j}  = \sum_{j \in J\setminus R}p_{jk_j} + \sum_{j \in J\cap R}p_{jk_j},$$

where $R = \bigcup_{i\in N} R_i$ is the set of items allocated in the local search. Note that each of the prices in the first sum is 0, by definition. The second sum can be bounded by,

$$ \sum_{j \in J\cap R}p_{jk_j} = \sum_{i \in \overline{N}} \sum_{j \in R_{i}}p_{jk_j} \leq \sum_{i \in \overline{N}} 1 = n$$ where in the first inequality we use \Cref{lemma:price-sum}.

Now we bound the sum of $|H_i|$:

$$\sum_{i \in \overline{N}}|H_{i}| + \sum_{i \in N\setminus \overline{N}}|H_{i}| = |H| = n.$$

Putting these two bounds gives us

$$\prod_{i \in N} \left(\frac{v_i(\OPT_i)}{v_i(T_i)}\right) \leq \left(6+\frac{4m}{n}\hat{\eps}\right)^{n} \leq 6^{n}(1+\frac{m}{n}{\hat\eps})^{n} \leq 6^n(1+\eps)^n$$

where we use Bernoulli's inequality in the last step.    
\end{proof}

We can now show the existence of the required matching $\sigma$.

\begin{restatable}[Existence of $\sigma$]{lemma}{InfeasibleToFeasible}
There exists a matching $\sigma: N \to H \cup \{\emptyset\}$ such that $\NSW(\cR,\sigma) \geq \frac{\NSW(\OPT)}{6(1+\eps)}$.
\label{lemma:matching}
\end{restatable}

We can construct the required matching $\sigma$ using the previously defined intermediate mapping $\cT$. To do so, we first restate a lemma about bipartite matchings given in \citep{GHL+23approximating}.

\begin{restatable}[\citet{GHL+23approximating}]{lemma}{bipartite}
    \label{lemma:bipartite}
Let $G = (X, Y; d)$ be a complete bipartite graph with edge weights $d\in (\RR \cup \{-\infty\})^{X\times Y}$. Let $\tau: X \to Y$ be a maximum-weight $X$-perfect matching, and let $Z \coloneqq \tau(X) \subseteq Y$ denote the set of matched nodes in $Y$. For some integer $k \geq 1$, let us obtain from $G$ the bipartite graph $G' = (X, Y'; d')$
by creating $k$ copies of every node in $Y \setminus Z$; thus, $|Y'| = k|Y| - (k-1)|X|$. Then, for any $X' \subseteq X$, there is a maximum-weight $X'$-perfect matching $\rho : X' \to Y'$ such that $\rho(X') \subseteq Z$.
\end{restatable}

\begin{proof} (of \Cref{lemma:matching})
Create a bipartite graph $G = (N, M; d)$ with weights $d_{ij} = \log v_i(j)$. Then, $\tau$ computed in the first phase of the algorithm is a max-weight $N$ perfect matching of $G$ which matches all items in $H$. Let $X' = N_{g} \cup N_{\ell}$ and create $k = n$ copies of $M\setminus H$. In this modified graph, we can create a $X'$-perfect matching by mapping $i\in N_{g}$ to $g(i)$ and $i \in N_{\ell}$ to a copy of $\ell(i)$. Note that this matching is the mapping $\cT$ restricted to $X'$. Using \Cref{lemma:bipartite}, we have the existence of a matching $\sigma : X' \to H$ with at least as much weight as that of the $X'$-perfect matching we defined. Extend this matching $\sigma$ to $N$ by defining $\sigma(i) = \emptyset$ for $i \in N_R$. Now, we can see that $\NSW(\cR, \sigma) \geq \NSW(\cT)$ and we have already shown $\NSW(\cT) \geq \frac{\NSW(\OPT)}{6(1+\eps)}$ in \Cref{lemma:intermediate}.
\end{proof}

We can now present the proof of \Cref{lemma:exact-capacitated-1-sided}.

\begin{proof} (of \Cref{lemma:exact-capacitated-1-sided})
    We first prove that \Cref{alg:NSW-template} terminates in polynomial time and makes a polynomial number of value queries. The first phase can be computed in polynomial time by finding a max-weight matching of the bipartite graph whose vertex sets are the set of agents $N$ and the set of items $M$. The edge $(i,j)$ between agent $i$ and item $j$ has weight $\log(v_i(j))$. This construction clearly requires only a polynomial number of value queries. The second phase terminates in polynomial time with a polynomial number of value queries as shown in \Cref{lemma:local}. Finally, the third phase also requires polynomial time and a polynomial number of value queries because we can create a bipartite graph as in the first phase where the weight of edge $(i,j)$ for an agent $i\in N$ and an item $j\in H$ is $\log(v_i(R_i + j))$.

    The algorithm returns a feasible partial allocation because $|R_i|$ is guaranteed to be $c_i-1$, and we allocate one more item to each agent in the final phase. Since our algorithm finds a matching that maximizes $\NSW(\cR, \delta)$, we get $\NSW(\ALG) \geq \frac{\NSW(\OPT)}{6(1+\eps)}$ using \Cref{lemma:matching}. To get the factor of $(6+\eps)$, we can scale down $\eps$ by a factor of $6$.
\end{proof}

To prove \Cref{theorem:capacitated-1-sided}, we will present a reduction from \CapacitatedOneSidedNash{} to \ExactCapacitatedOneSidedNash{}.

\begin{proof} (of \Cref{theorem:capacitated-1-sided})
    Let $\I =\langle N, M, \V, C \rangle$ be the given instance of \CapacitatedOneSidedNash{}. We create an instance $\I' = \langle N, M', \V', C\rangle$ of \ExactCapacitatedOneSidedNash{} as follows: If $|M| \geq \sum_{i \in N} c_i$, then set $M' = M$. Otherwise, if $|M| < \sum_{i \in N} c_i$, we create $M'$ by adding dummy items to $M$ till $|M'| = \sum_{i\in N} c_i$. These dummy items provide $0$ value to all agents. Formally, $v'_i(S) = v_i(S\cap M)$ for all $i\in N$ and $S\subseteq M'$. 

    Let $\OPT$ be an optimal partial allocation of $\I$. We can convert this partial allocation to a feasible partial allocation for $\I'$ by allocating some dummy items to agents until their exact capacity constraint is reached. Note that the $\NSW$ of $\OPT$ remains unchanged in this process because dummy items do not provide any value. Next, let $\OPT'$ be an optimal partial allocation of $\I'$. We can convert this partial allocation to a feasible allocation of $\I$ by removing all the allocated dummy items. Again, the $\NSW$ of $\OPT'$ is unchanged. Hence, we get that $\NSW(\OPT) = \NSW(\OPT')$.

    We know from \Cref{lemma:exact-capacitated-1-sided} that, given as input the instance $\I'$, \Cref{alg:NSW-template} returns a partial allocation $A'$ that satisfies $\NSW(A') \geq \frac{\NSW(\OPT')}{(6+\eps)}$. After removing all dummy items from this partial allocation, we get a feasible partial allocation of $\I$, say $A$. 
    
    Since $\NSW(\OPT) = \NSW(\OPT')$ and $\NSW(A) = \NSW(A')$, we get that $\NSW(A) \geq \frac{\NSW(\OPT)}{(6 + \eps)}$.
\end{proof}

\section{Approximation Algorithms for the Two-Sided Model}
\label{sec:TwoSidedAlgo}

We will now present our algorithmic results for two-sided Nash welfare. Our main result is a $1.33$-approximation algorithm for \CapacitatedTwoSidedNash{} under subadditive valuations~(\Cref{theorem:capacitated-2-sided}). The running time of the algorithm is strongly polynomial in $n$ and $m$, i.e., it doesn't depend on the size of the valuations. This result significantly improves upon an existing $\sqrt{\texttt{OPT}}$-approximation algorithm for positive additive valuations~\citep{JV24maximizing}.

A corollary of our main result is that when the number of firms is constant, we obtain a polynomial-time (and polynomial-query) approximation scheme (\PTAS{}) for the two-sided problem under subadditive valuations~(\Cref{cor:PTAS-TwoSided}). This result improves upon a quasipolynomial-time approximation scheme (\QPTAS{}) for a constant number of firms under polynomially-bounded additive valuations~\citep{JV24maximizing}. 
We will start by proving our main result in \Cref{theorem:capacitated-2-sided}.

\begin{restatable}[]{theorem}{TwoSidedApprox}
There exists a $1.33$-approximation algorithm for \CapacitatedTwoSidedNash{} under subadditive valuations. The algorithm runs in $\poly(n,m)$ time and makes $\poly(n,m)$ number of value queries.
\label{theorem:capacitated-2-sided}
\end{restatable}

Towards proving \Cref{theorem:capacitated-2-sided}, we will first state a lemma that provides an upper bound on the Nash welfare of any many-to-one matching. The bound is in terms of the firms' values for only their \emph{favorite} matched worker.

\begin{lemma}
    Let $\I = \langle F, W, \V, \W, C \rangle$ be an instance of \CapacitatedTwoSidedNash{} with subadditive valuations and let $\mu$ be any feasible many-to-one matching for $\I$. Let $j_{i} \in \argmax_{j \in \mu_i} v_i(j)$ be firm $i$'s favourite worker in $\mu_i$. Then, %
    \begin{center}
        $\NSW(\mu) \leq 1.33\left(\prod_{i\in F} v_i(j_i) \prod_{j \in W } w_j(\mu_j)\right)^{\frac{1}{m+n}}.$
    \end{center}
\label{lemma:favourite-worker}
\end{lemma}

\begin{proof} (of \Cref{lemma:favourite-worker})
We know that among the workers in $\mu_i$, firm $i$ assigns the highest value to the worker $j_{i}$. By the subadditivity of $v_i$, we get that $v_i(\mu_i) \leq |\mu_i| v_i(j_{i})$.

Substituting this inequality in the expression of $\NSW(\mu)$ gives:
\begin{equation} \label{eq:2-sided-nsw-ub}
    \NSW(\mu) \leq \left(\prod_{i\in F} |\mu_i| \prod_{i\in F} v_i(j_i) \prod_{j \in W } w_j(\mu_j)\right)^{\frac{1}{m+n}}.
\end{equation}
Since $\sum_{i \in F} |\mu_i| \leq m $, using AM-GM inequality, we have $\prod_{i \in F} |\mu_i| \leq \left(\frac{m}{n}\right)^n$. Hence, $ \left(\prod_{i\in F} |\mu_i| \right)^{\nicefrac{1}{m+n}} \leq \left(\frac{m}{n}\right)^{\nicefrac{n}{m+n}} = x^{\nicefrac{1}{1+x}}$ , where $x = \nicefrac{m}{n}$. The function $x^{\nicefrac{1}{1+x}}$ is upper bounded by $1.3211$  for all $x > 0$ . Substituting this bound in \Cref{eq:2-sided-nsw-ub} completes the proof.
\end{proof}

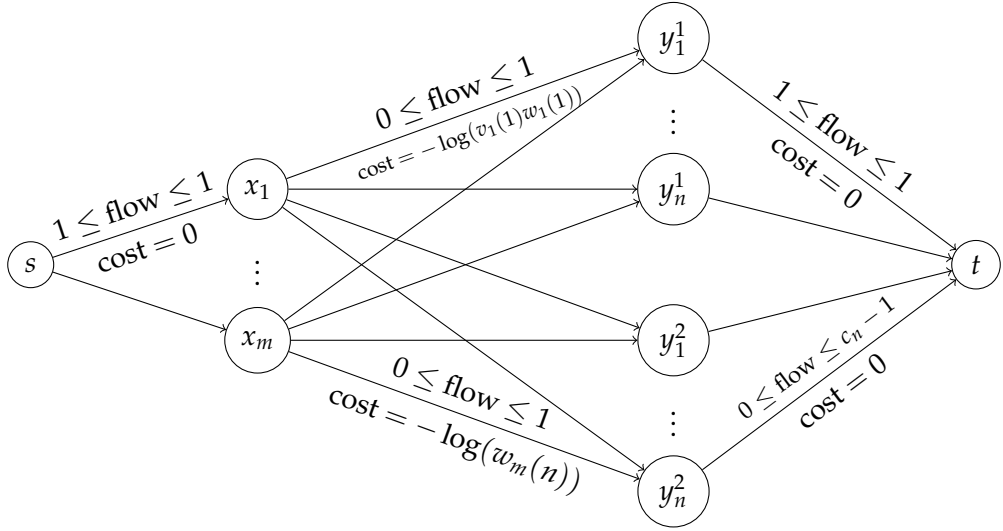
\begin{figure}[t]
    \centering
    \resizebox{0.8\columnwidth}{!}{\input{fig}}
    \caption{\MinCostFLow{} network used in the proof of \Cref{theorem:capacitated-2-sided}. %
    The edge labels show the cost per unit flow and the lower and upper bounds on the flow.}
    \label{fig:flow}
\end{figure}

With this bound on $\NSW$ in place, we will now create a \MinCostFLow{} instance. %
Recall that in the \MinCostFLow{} problem, we are given a flow network where each edge has a lower as well as an upper bound on the flow that can pass through it, and a cost per unit flow associated with it. The objective is to find a feasible flow with the minimum total cost. The problem is known to be solvable in strongly polynomial time with a simple algorithm that starts with a feasible flow and repeatedly cancels a residual cycle with negative cost, specifically, one with the minimum mean cost~\citep{GT89finding}.

\begin{remark}
    It is relevant to note that although the costs of edges in our flow network are specified in terms of logarithms, it is not essential to \emph{compute} the log values. Indeed, the algorithm of~\citet{GT89finding} requires computing a cycle with the minimum mean cost. Such a cycle can be computed in strongly polynomial time~\citep{K78characterization,CM17note}. This algorithm only requires computing the \emph{sum} of edge costs along a walk of a certain length, which amounts to computing the \emph{product} of the corresponding valuations. The latter computation can be done in $\O(nm)$ time under the standard arithmetic model (see~\Cref{sec:Preliminaries}). This observation allows us to circumvent the explicit calculation of the (possibly irrational) log value and facilitates a strongly polynomial running time.
\end{remark}

\begin{proof} (of \Cref{theorem:capacitated-2-sided})
Let $\I = \langle F, W, \V, \W, C \rangle$ be the given instance of \CapacitatedTwoSidedNash{}. For an edge $e$ in a flow network, we will use $L(e)$ and $U(e)$ to denote the lower and the upper bounds on the flow through $e$, respectively, and will write $\cost(e)$ to denote the cost per unit flow through $e$. Also, we will use $f(e)$ to denote the flow through the edge $e$. The flow network is constructed as follows (see \Cref{fig:flow}):
\begin{itemize}%
    \item Create $m$ vertices $x_1, \dots, x_m$ corresponding to the $m$ workers.
    \item Create $2n$ vertices $y^1_1, \dots, y^1_n$ and $y^2_1, \dots, y^2_n$ corresponding to the $n$ firms. We call $y^1_i$ the \emph{main copy} and $y^2_i$ the \emph{secondary copy} of firm $i$.
    \item Create a source vertex $s$. For each worker $j$, add an edge from $s$ to $x_j$ with $L(s\rightarrow x_j) = 1, U(s\rightarrow x_j) = 1$, and $\cost(s\rightarrow x_j) = 0$.
    \item Create a sink vertex $t$. For each firm $i$, add an edge from $y_i^1$ to $t$ with $L(y_i^1 \rightarrow t)= 1$, $U(y_i^1 \rightarrow t) = 1$ and $\cost(y_i^1 \rightarrow t) = 0$. Also, add an edge from $y_i^2$ to $t$ with $L(y_i^2 \rightarrow t)= 0$, $U(y_i^2 \rightarrow t) = c_i - 1$ and $\cost(y_i^2 \rightarrow t) = 0$.
    \item For all $j\in W$ and $i\in F$,  such that $v_i(j) \neq 0$ and $w_j(i) \neq 0$, add an edge from $x_j$ to $y^1_i$ with $L(x_j \rightarrow y_i^1) = 0, U(x_j \rightarrow y_i^1) = 1$  and $\cost(x_j \rightarrow y_i^1) = - \log(v_i(j) w_j(i)) $.
    \item For all $j\in W$ and $i\in F$ such that $w_j(i) \neq 0$, add an edge from $x_j$ to $y^2_i$ with $L(x_j \rightarrow y_i^2) = 0, U(x_j \rightarrow y_i^2) = 1$  and $\cost(x_j \rightarrow y_i^2) = - \log(w_j(i)) $. 

\end{itemize}

Let $\mu^\star$ be a Nash optimal many-to-one matching of $\I$. We will assume that, under $\mu^\star$, every firm is assigned at least one worker for whom it has a nonzero value. Likewise, each worker is matched to a firm that it values positively. Indeed, if this were not the case, then the optimal Nash welfare would be zero, making every possible feasible assignment compliant with the approximation value.
 For a firm $i$, let $j^\star_i$ denote its favorite assigned worker under $\mu^\star$. Notice that the following flow vector $f$ is feasible in the above network:
\begin{itemize}%
\item $f(s \rightarrow x_j) = 1$ for all the workers $j$.
\item $f(y_i^1 \rightarrow t) = 1$ for all firms $i$.
    \item $f(x_{j^\star_i }\rightarrow y_i^1) = 1$ for all firms $i$.
    \item $f(x_j \rightarrow y_i^2) = 1$ for all $j \in \mu^\star_i \setminus \{j^\star_i\}$ 
    \item $f(y_i^2 \rightarrow t) = |\mu^\star_i| - 1$ for all firms $i$.
\end{itemize}

For all other edges $e$ in the network, we set $f(e) = 0$. Clearly, the cost of $f$ is:
\begin{equation} \label{eq:cost_f}
     \cost(f) = - \sum_{i \in F} \log(v_i(j^\star_i)) - \sum_{j \in W} \log(w_j(\mu^\star_j)).
\end{equation}

Our algorithm finds the minimum-cost flow, say $f^\star$, in the network. Note that any feasible flow from $s$ has the same value as the maximum flow. Furthermore, since the upper and lower bounds on the flow are integral, we have that the resulting flow is integral. Thus, the flow $f^\star$ induces a valid many-to-one matching, say $\bar{\mu}$, of workers to firms. Under $\bar{\mu}$, each worker $j$ is assigned to a unique firm $i$ (i.e., $\bar{\mu}_j \coloneqq i$) such that $f^\star(x_j \rightarrow y_i^1) = 1$ or $f^\star(x_j \rightarrow y_i^2) = 1$. For each firm $i$, let $\bar{\mu}_i$ be the the set of workers assigned to the firm $i$, and let $\bar{j}_i$ be the unique worker $j$  for which $f^\star(x_j \rightarrow y_i^1) = 1$. Therefore, the cost of $f^\star$ is:
\begin{equation} 
      \cost(f^\star) = - \sum_{i \in F} \log(v_i(\bar{j}_i)) - \sum_{j \in W} \log(w_j(\bar{\mu}_j)).
\label{eq:cost_f_star}
\end{equation}

Using the cost optimality of $f^\star$ along with \Cref{eq:cost_f_star,eq:cost_f},
\begin{equation} \label{eq:cost_optimality}
      \prod_{i \in F} v_i(\bar{j}_i) \prod_{j \in W} w_j(\bar{\mu}_j) \geq \prod_{i \in F} v_i(j^\star_i) \prod_{j \in W} w_j(\mu^\star_j).
\end{equation}

Therefore, the Nash welfare of our assignment is at least:
\begin{align*} 
    \NSW(\bar{\mu}) &=   \left(\prod_{i \in F} v_i(\bar{\mu}_i) \prod_{j \in W} w_j(\bar{\mu}_j) \right) ^{\frac{1}{m+n}} \\
    &   \geq \left(\prod_{i \in F} v_i(\bar{j}_i) \prod_{j \in W} w_j(\bar{\mu}_j) \right) ^{\frac{1}{m+n}}\\
    &   \geq \left(\prod_{i \in F} v_i(j^\star_i) \prod_{j \in W} w_j(\mu^\star_j)\right)^{\frac{1}{m+n}} \\
    &   \geq \frac{\NSW(\mu^\star)}{1.33},
\end{align*}
where the second-last inequality follow from \Cref{eq:cost_optimality} and the last inequality follows from \Cref{lemma:favourite-worker}. This finishes the proof of \Cref{theorem:capacitated-2-sided}.
\end{proof}

We will now show that when the number of firms is constant, the algorithm in the proof of \Cref{theorem:capacitated-2-sided} can be used to obtain a polynomial-time approximation scheme (\PTAS{}) for the two-sided problem under subadditive valuations.

\begin{restatable}[PTAS for constant number of firms]{corollary}{PTAS}
For any $\eps > 0$ and a constant number of firms, there exists a $(1+\eps)$-approximation algorithm for \CapacitatedTwoSidedNash{} under subadditive valuations. The algorithm makes $\poly(m)$ number of value queries and runs in $\poly(m)$ time.
\label{cor:PTAS-TwoSided}
\end{restatable}
\begin{proof} (of \Cref{cor:PTAS-TwoSided})
    If $\eps \geq 0.33$, the algorithm in \Cref{theorem:capacitated-2-sided} is already a $(1+\eps)-$ approximate algorithm. Hence, we assume $\eps < 0.33$. Our algorithm runs the \MinCostFLow{} algorithm of \Cref{theorem:capacitated-2-sided} if $m \geq \frac{n}{\eps^2}$, and otherwise solves the problem exactly by iterating over all the $O(n^m)$ many-to-one matchings. The running time of our algorithm is $O\left(n^{\nicefrac{n}{\eps^2}} \cdot \text{poly}(n, m)\right)$.
    
    The approximation ratio of our algorithm, as derived from the proof of \Cref{lemma:favourite-worker}, equals $r = x^{\nicefrac{1}{1 + x}}$, where $x = \frac{m}{n} > \frac{1}{\eps^2} > \frac{1}{0.33^2} > 9$. It can be verified that $\frac{\log(x)}{1 + x} < \frac{1}{x^{0.75}}$ for all $x > 9$, and that $\eps^{1.5} < \log(1 + \eps)$ for all $\eps < 0.33$. Therefore, $\log (r) = \frac{\log(x)}{1 + x} < \frac{1}{x^{0.75}} < \eps^{1.5} < \log(1 + \eps)$, as desired for the $(1+\eps)$-approximation.
\end{proof}

\section{Weighted Nash welfare under capacity constraints}
In this section, we modify the algorithm of \cite{feng_et_al:LIPIcs.ICALP.2024.63} to obtain an $e^{1/e + \epsilon}-$ approximation algorithm for maximizing weighted Nash welfare in the two sided model under capacity constraints, with firms having additive valuations. Here, firm $i$ has a weight $\eta_i$ and each worker $j$ has a weight $\zeta_j$, such that $\sum_{i \in F} \eta_i + \sum_{j \in W} \zeta_j = 1$. The weighted Nash welfare for a many to one matching $\mu$ is defined as:

\begin{equation*}
    \NSW(\mu) = \prod_{i \in F} v_i(\mu_i)^{\eta_i} \prod_{j \in W} w_j(\mu_j)^{\zeta_j}
\end{equation*}

Notably, the above definition incorporates the one sided model if we put $\zeta_j = 0$ for all $j \in W$. In the rest of the section, we will prove the following theorem:

\begin{theorem} \label{thm:approx_guarantee}
    For any $\epsilon > 0$, there exists a polynomial time $e^{(\sum_{i \in F} \eta_i)/e + \epsilon}$-approximation algorithm for the weighted \CapacitatedTwoSidedNash{} under additive valuations.
\end{theorem}

    We begin with a natural modification of the configuration LP from \cite{feng_et_al:LIPIcs.ICALP.2024.63}, with the changes highlighted in red:

\begin{align}
    \max \sum_{i \in F, S \subseteq  W} \left( \eta_i \cdot \ln v_i(S) \textcolor{red}{+ \sum_{j \in S} \zeta_j \cdot \ln w_j(i)}\right) \cdot y_{i, S}  \qquad \text{s.t.} \tag{Conf-LP}\label{CLP}
\end{align} \vspace*{-10pt}
\begin{align}
    \sum_{i \in F, S \ni j} y_{i, S} &\leq 1 &\quad &\forall j \in W \label{LPC:j-assigned-once}\\
    \sum_{S \subseteq W} y_{i, S} &= 1 &\quad &\forall i \in F \label{LPC:i-gets-one-set}\\
    \textcolor{red}{y_{i,S}} &\textcolor{red}{= 0} &\quad &\textcolor{red}{\forall i \in F, S \subseteq W: |S| > c_i}\\
    y_{i, S} &\geq 0 &\quad &\forall i \in F, S\subseteq W
\end{align}

Here, we have a variable $y_{i,S}$ corresponding to each firm $i$ and each subset $S \subseteq W$ of workers. In the integer version of the LP, the variable $y_{i,S}$ is equal to 1 if firm $i$ is assigned the set of workers $S$, and 0 otherwise. The first constraint ensures that each worker is assigned to at most one firm, while the second constraint ensures that each firm is assigned exactly one set of workers. The third constraint ensures that the number of workers assigned to a firm does not exceed its capacity. The last constraint ensures that all variables are non-negative.

We solve the (~\ref{CLP}) within an additive error of $\ln(1 + \epsilon)$ using the same method as in \cite{feng_et_al:LIPIcs.ICALP.2024.63}. We design an approximate separation oracle for the dual and identify a polynomial number of variables $y_{i,S}$, such that the rest of the variables can be set to $0$ in (~\ref{CLP}) without significantly changing the optimal value. We defer the details to ~\Cref{sec:solving-configuration-lp}. 

\subsection{Rounding and its analysis}
We use the rounding algoithm of \cite{feng_et_al:LIPIcs.ICALP.2024.63} without any modifications. Let $\mu$ be the final many to one matching returned by the rounding algorithm. For a firm $i$, let $\mu_i$ be the set of workers assigned to firm $i$ under $\mu$. Also, for a worker $j$, let $\mu_j$ be the firm that that $j$ is assigned to, under $\mu$.
We will use the following properties of their rounding algorithm in our analysis:
\begin{enumerate}
    \item Every agent $i \in F, |\mu_i| \leq \lceil \sum_{j\in W} x_{i,j}\rceil$, where $x_{i,j}$ equals $\sum_{S \ni j} y_{i,S}$.
    \item A worker $j \in W$ is assigned to a firm $i$ with probability exactly $x_{i,j} = \sum_{S \ni j} y_{i,S}$.
\end{enumerate}

From the first property, we have:

\begin{align*}
    |\mu_i| &\leq \lceil \sum_{j \in W} x_{i,j}\rceil
    = \lceil \sum_{j \in W} \sum_{S \ni j} y_{i,S} \rceil
    = \lceil\sum_{S \subseteq W} y_{i,S} \cdot |S| \rceil
    \leq \lceil \sum_{S \subseteq W} y_{i,S} \cdot c_i \rceil
    = c_i
\end{align*}

Hence, the capacity constraints are satisfied in the final allocation. The expected log of the utility of the firms is bounded in \cite{feng_et_al:LIPIcs.ICALP.2024.63} as:

\begin{lemma}(Lemma 4 in \cite{feng_et_al:LIPIcs.ICALP.2024.63}) \label{lem:expected-utility-firms}
    For every $i \in F$, we have:
    $$\mathbb{E}[\ln(v_i(\mu_i))] \geq \sum_{S \subseteq W} y_{i,S} \ln v_i(S) - \frac{1}{e}$$
    
\end{lemma}

Also, the following lemma follows directly from the second property:

\begin{lemma} \label{lem:expected-utility-workers}
    For every $j \in W$, we have:
    $$\mathbb{E}[\ln(w_j(\mu_j))] = \sum_{i \in F}\sum_{S \ni j} y_{i,S} \ln w_j(i)$$
\end{lemma}

Let $\mu^\star$ denote the (weighted) nash optimal many to one matching. Combining ~\Cref{lem:expected-utility-firms,lem:expected-utility-workers}, we have:

\begin{align*}
    \mathbb{E}[\ln \NSW(\mu)] &= \sum_{i \in F} \eta_i \mathbb{E}[\ln(v_i(\mu_i)))] + \sum_{j \in W} \zeta_j \mathbb{E}(\ln(w_j(\mu_j)))\\
    & \geq \sum_{i \in F} \eta_i \left( \sum_{S \subseteq W} y_{i,S} \ln v_i(S) - \frac{1}{e} \right) + \sum_{j \in W} \zeta_j\sum_{i \in F}\sum_{S \ni j} y_{i,S} \ln w_j(i)\\
    &= \sum_{i \in F, S \subseteq  W} \left( \eta_i \cdot \ln v_i(S) + \sum_{j \in S} \zeta_j \cdot \ln w_j(i)\right) \cdot y_{i, S} - \frac{\sum_{i \in F} \eta_i}{e}\\
    & \geq \ln \NSW(\mu^\star) - \epsilon - \frac{\sum_{i \in F} \eta_i}{e}
\end{align*}

The final inequality has an $\epsilon$ term, because we solve the (\ref{CLP}) within an additive error of $\ln(1 + \epsilon) \leq \epsilon$. Now, applying Jensen's inequality proves ~\Cref{thm:approx_guarantee}.

\subsection{An improved algorithm for unweighted \CapacitatedTwoSidedNash with additive valuations}
Note that, in ~\Cref{sec:TwoSidedAlgo}, we obtain an $ x^{\frac{1}{x+1}}-$ approximation algorithm for the unweighted \CapacitatedTwoSidedNash problem, where $x = \frac{m}{n}$. Also, substituting $\eta_i = \frac{1}{m + n}$ for all $i \in F$ and $\zeta_j = \frac{1}{m + n}$ for all $j \in W$, we get an $e^{\frac{m}{e(m+n)} + \epsilon} = e^{\nicefrac{1}{e(x + 1)} + \epsilon}$ approximation algorithm.

The expression $\min(x^{\nicefrac{1}{(x+1)}}, e^{\nicefrac{1}{e(x+1)}})$ is maximized at $x = e^{1/e}$. Hence, there exists a polynomial time $e^{\nicefrac{1}{e(e^{1/e}+1)} + \epsilon} \approx (1.163 )-$ approximation algorithm for the unweighted \CapacitatedTwoSidedNash problem with additive valuations.

\section{Hardness Results}

\label{sec:HardnessResults}

In this section, we will show that the problem of maximizing two-sided Nash social welfare is \APXh{}; specifically, the problem is \NPh{} to approximate within a factor of $1.0000759$. This result holds even in the absence of capacity constraints and even under additive valuations. Prior to our result, only \NPh{}ness was known for this problem~\citep{JV24maximizing}.

\begin{restatable}[Hardness for two-sided Nash welfare]{theorem}{} 
Unless $\P{}=\NP{}$, no polynomial-time algorithm can approximate \TwoSidedNash{} to within a factor of $1.0000759$ even under additive valuations.%
\label{thm:Two_Sided_Hardness_Nash}
\end{restatable}
To prove \Cref{thm:Two_Sided_Hardness_Nash}, we will use~\Cref{lem:one_sided_hardness_linear_items}, as stated below. This lemma is based on a result of~\citet{GM21hardness}, who showed \APXh{}ness of \OneSidedNash{} under additive valuations. The key property from their reduction, needed in \Cref{lem:one_sided_hardness_linear_items}, is that the number of items in the reduced instance is at most a constant (specifically, $1.5$) times the number of agents.
\begin{lemma}[modified from ~\citep{GM21hardness}]
Unless $\P{} = \NP{}$, no polynomial-time algorithm can approximate \OneSidedNash{} with additive valuations to a factor smaller than $1.00019$, even when the number of items is at most $1.5$ times the number of agents (i.e., $m \leq 1.5 n$).
\label{lem:one_sided_hardness_linear_items}
\end{lemma}
We will now prove the \APXh{}ness of two-sided Nash welfare even in the absence of capacity constraints.
\begin{proof} (of \Cref{thm:Two_Sided_Hardness_Nash})
Consider an instance of \OneSidedNash{} instance denoted by $\I_1 = \langle N,M,\V_1,C \rangle$, where the number of items $m = |M|$ is at most $1.5$ times the number of agents $n = |N|$, and the valuations are additive.

We will create an instance $\I_2 = \langle F,W,\mathcal{V}_2,\mathcal{W},C \rangle$ of \TwoSidedNash{} as follows: The set of firms (respectively, workers) corresponds to the set of agents (respectively, items) in the one-sided instance $\I_1$, i.e., $F = N$ and $W = M$. A firm's valuations for the workers and its capacity are identical to the corresponding agent's valuations for the items and capacity, respectively; thus, $\mathcal{V}_2 = \mathcal{V}_1$. Finally, the workers have uniform valuations over the firms, i.e., for each worker $j \in W$ and each firm $i \in F$, $w_j(i) = 1$.

Note that, any allocation $A^{\I_1}$ of the items to the agents in the one-sided instance $\I_1$ corresponds naturally to a many-to-one matching $\mu^{\I_2}$ in the instance $\I_2$ and vice-versa. Here, the set of workers $\mu^{\I_2}_i$ assigned to firm $i$ under $\mu^{\I_2}$ is given by $ A^{\I_1}_i$, namely, the set of items allocated to agent $i$ in the allocation $A^{\I_1}$. Since the workers' valuations are uniformly equal to $1$, we have:
\[
\NSW(A^{\I_1}) = \left(\NSW(\mu^{\I_2})\right)^{\frac{m+n}{n}} = \left(\NSW(\mu^{\I_2})\right)^{1 + \frac{m}{n}}.
\]
Hence, any $\gamma$-approximate solution for the two-sided instance $\I_2$ yields a $ \left(\gamma^{1 + \nicefrac{m}{n}}\right)$- approximate solution for one-sided instance $\I_1$.  Therefore, using~\Cref{lem:one_sided_hardness_linear_items}, we have that unless $\P{}=\NP{}$, no polynomial-time algorithm can approximates \TwoSidedNash{} to a factor smaller than $\left( 1.00019 \right)^{\nicefrac{1}{1+1.5}} > 1.0000759$.
\end{proof}

\begin{remark}
It follows from the proof of \Cref{thm:Two_Sided_Hardness_Nash} that any reduction showing that \OneSidedNash{} is hard to approximate within a factor of $\alpha$ under additive valuations, for instances where $m \leq c n$ for some constant $c$, yields a $\alpha^{\nicefrac{1}{1+c}}$ hardness-of-approximation ratio for \TwoSidedNash{}.
\end{remark}

\section{Concluding Remarks}

We studied algorithmic aspects of maximizing Nash social welfare for one-sided and two-sided preferences under capacity constraints. We developed constant-factor approximation algorithms for both settings, complementing the \APXh{}ness results. Our algorithm for the one-sided problem provides the first constant-factor approximation algorithm for Nash welfare under submodular valuations and capacity constraints, while our result for the two-sided problem applies to subadditive valuations and significantly improves upon the existing $\sqrt{\texttt{OPT}}$-approximation for additive valuations.%

Our work opens up several directions for future work. Firstly, it would be interesting to obtain tight bounds for the approximations mentioned in \Cref{tab:Summary}. In particular, our hardness reduction in the two-sided setting uniformly assigns a valuation of $1$ to all workers, thereby essentially leveraging the hardness of the one-sided problem. An important question to address is whether stronger hardness-of-approximation results are attainable when workers' valuations are not uniform. Secondly, exploring extensions to more general constraints than capacities, such as matroid constraints, would be a fruitful direction for further investigation. Lastly, it would be interesting to explore if, for some choice of weights, there exists an axiomatic justification of the weighted Nash welfare objective in the two-sided setting.

\section*{Acknowledgments}
RV acknowledges support from SERB grant no. CRG/2022/002621, DST INSPIRE grant no. DST/INSPIRE/04/2020/000107, and iHub Anubhuti IIITD Foundation. JY acknowledges support from Google PhD Fellowship.

\bibliographystyle{plainnat} 
\bibliography{ms}

\clearpage

\appendix

\section{Additional Details}
\label{sec:appendix}
\subsection{Solving the configuration LP with small additive error} \label{sec:solving-configuration-lp}

Note that, the utility of each firm is at most $m$ times the utility of the firm for the top worker assigned to it. Hence, our algorithm from ~\Cref{sec:TwoSidedAlgo} is an $m$-approximation to the optimal weighted nash welfare, and hence is a $O(m)$ approximation to the optimum value of (~\ref{CLP}) (note that the analysis of the rounding algorithm proves that the integrality gap is at most $e^{1/e}$). Hence, we can assume that in $O(\log(\frac{m}{\epsilon}))$ guesses, we have a number $o$, such that the optimum value of (~\ref{CLP}) is in the range $(o, o + \epsilon/3]$. We replace the objective by a constraint in the dual to obtain the following {\bf infeasible} LP:
\begin{align}
    \tag{Dual-LP}\label{DLP}
    \sum_{j \in W} \alpha_j + \sum_{i \in F} \beta_i \leq o
\end{align} \vspace*{-10pt}
\begin{align}
    \sum_{j \in S} \alpha_j + \beta_i &\geq \eta_i \cdot \ln v_i(S) \textcolor{red}{+ \sum_{j \in S} \zeta_j \cdot \ln w_j(i)} &\quad &\forall i \in F, S \subseteq W\textcolor{red}{: |S| \leq c_i} \label{LDC:i-gets-set}\\
    \alpha_j &\geq 0 &\quad &\forall j \in W \label{LDC:j-non-neg}
\end{align}

We know that the above LP is infeasible. Now, we design an approximate separation oracle for the dual LP. Suppose we are given some dual variables $ \alpha \in \mathbb{R}_{\geq 0}^W$ and $\beta \in \mathbb{R}^F$ such that $\sum_{j \in W} \alpha_j + \sum_{i \in F} \beta_i \leq o$. Our approximate separation oracle will return a set $S' \subseteq W$ and a firm $i' \in F$ such that $|S'| \leq c_{i'}$ and:
\begin{align}
\sum_{j \in S'} \alpha_j + \beta_{i'} < \eta_{i'} \cdot \ln (v_{i'}(S') (1 + \epsilon / 2)) + \sum_{j \in S'} \zeta_j \cdot \ln w_j(i') \label{eq:separation}
\end{align}

Let $i$ be a firm and $S$ be a subset of workers for which ~\Cref{LDC:i-gets-set} is violated (such a pair $(i,S)$ exists because (~\ref{DLP}) is infeasible and $\sum_{j \in W} \alpha_j + \sum_{i \in F} \beta_i \leq o$). We will iterate over all $n$ possibilities of $i$. Then, if $\eta_i = 0$, we will sort the workers according to $\alpha_j - \zeta_j \ln w_j(i)$ (assume $\ln 0 = -\infty$), and test ~\Cref{eq:separation} for all prefixes of length upto $c_i$ of the sorted list. Otherwise, let us define $\alpha'_j  = \frac{1}{\eta_i} ( \alpha_j - \zeta_j \ln w_j(i))$ for all workers $j \in W$. Then, we are interested in finding a set $S$ such that $\sum_{j \in S} \alpha'_j + \beta_i/\eta_i < \ln (v_i(S) (1 + \epsilon / 2))$. For this purpose, we employ the standard dynamic programming based approximate solution for the knapsack problem. We iterate over all the $m$ possibilities of $j^\star \in S$ with the largest $v_i(\cdot)$ value, and throw away all $j'$ with $v_i(j') > v_i(j^\star)$. Then, let us define $\hat{v}_i(j) = \lfloor \frac{2m v_{i}(j)}{\eps v_{i}(j^\star)} \rfloor$. Then, using dynamic programming, we find, for all integers $0 \le x \le m, 0 \le y \le \frac{2m^2}{\epsilon}$, the smallest possible value of $\sum_{j \in S'} \alpha'_j$ such that $j^\star \in S', |S'| = x$ and $\sum_{j \in S'} \hat{v}_i(j) = y$. In $O(m^3/\epsilon)$ guesses, we can guess the value of $|S|$ and $\sum_{j \in S} \hat{v}_i(j)$. Hence, we can find a set $S'$ with $|S'| = |S|$, $\sum_{j \in S'} \hat{v}_i(j) = \sum_{j \in S} \hat{v}_i(j)$, and 

$$\sum_{j \in S'} \alpha'_j + \frac{\beta_i}{\eta_i} \leq \sum_{j \in S} \alpha'_j + \frac{\beta_i}{\eta_i} < \ln v_i(S) \leq \ln( v_i(S') (1 + \epsilon/2)).$$ 

Here, the first inequality follows from the correctness of the dynamic program, the second inequality follows from the violation of ~\Cref{LDC:i-gets-set} for $(i,S)$, and the last inequality follows from:

$$\hat{v}_i(j) \cdot \frac{\epsilon v_i(j^{\star})}{2m} \leq v_i(j) < (\hat{v}_i(j) + 1) \cdot \frac{\epsilon v_i(j^{\star})}{2m}.$$

Hence, for any constant $\epsilon > 0$, using ellipsoid method, we can find a polynomial number of inequations of the form $\sum_{j \in S'} \alpha_j + \beta_i \geq \eta_i \cdot \ln (v_i(S') (1 + \epsilon / 2)) + \sum_{j \in S'} \zeta_j \cdot \ln w_j(i)$, that are infeasible to all be satisfied along with $\sum_{j \in W} \alpha_j + \sum_{i \in F} \beta_i \leq o$. Therefore, consider (~\ref{CLP}) with all the $v_i(\cdot)$ values scaled by $1 + \epsilon / 2$, subject to all $y_{i,S} = 0$ for $i, S$ not corresponding to the infeasible inequalities. The optimum value of this new LP is at least $o$. Therefore, we can find a solution $y$ with:

\begin{align}
    &\sum_{i \in F, S \subseteq  W} \left( \eta_i \cdot \ln v_i(S) + \sum_{j \in S} \zeta_j \cdot \ln w_j(i)\right) \cdot y_{i, S}\\
    &=\sum_{i \in F, S \subseteq  W} \left( \eta_i \cdot \ln (v_i(S) (1 + \epsilon / 2)) + \sum_{j \in S} \zeta_j \cdot \ln w_j(i)\right) \cdot y_{i, S} - \sum_{i \in F, S \subseteq W} \eta_i \cdot \ln(1 + \epsilon/2) \cdot y_{i,S} \\
    &\geq o - \sum_{i \in F, S \subseteq W} \eta_i \cdot \ln(1 + \epsilon/2) \cdot y_{i,S} \\
    &= o - \sum_{i \in F} \eta_i \ln(1 + \epsilon/2) \geq o - \ln(1 + \epsilon / 2) \\
\end{align}

Since the optimal value of (~\ref{CLP}) is at most $o + \epsilon / 3$, we get a solution $y$ with additive error at most $\ln(1+\epsilon)$ for small enough $\epsilon$.

\end{document}

%% file: fig.tex
\begin{tikzpicture}
	\begin{pgfonlayer}{nodelayer}
\node [style={std_node}] (0) at (-4.5, 3) {$x_1$};
		\node [style={std_node}] (2) at (-4.5, 1) {$x_m$};
		\node [style=none] (9) at (-4.5, 2) {$\vdots$};
		\node [style={std_node}] (10) at (1, 5) {$y^1_1$};
		\node [style={std_node}] (11) at (1, 3) {$y^1_n$};
		\node [style=none] (12) at (1, 4) {$\vdots$};
		\node [style={std_node}] (13) at (1, 1) {$y^2_1$};
		\node [style={std_node}] (14) at (1, -1) {$y^2_n$};
		\node [style=none] (15) at (1, 0) {$\vdots$};
		\node [style={std_node}] (16) at (-7.5, 2) {$s$};
		\node [style={std_node}] (17) at (5, 2) {$t$};
	\end{pgfonlayer}
	\begin{pgfonlayer}{edgelayer}
		 \draw[->] (0) -- node[above, sloped] { $0 \leq $ flow $\leq 1$} node[below, sloped] { \scriptsize cost $= -\log(v_1(1)w_1(1))$} (10);
		\draw[->] (0) to (11);
		\draw[->] (0) to (13);
		\draw[->] (0) to (14);
		\draw[->] (2) to (10);
		\draw[->] (2) to (11);
		\draw[->] (2) to (13);
	 \draw[->] (2) -- node[above, sloped] {$0 \leq $ flow $\leq 1$} node[below, sloped] {  cost $= -\log(w_m(n))$} (14);
	 \draw[->] (16) -- node[above, sloped] { $1 \leq $ flow $\leq 1$} node[below, sloped] {  cost $= 0$} (0);
		\draw[->] (16) to (2);
	 \draw[->] (10) -- node[above, sloped] { $1 \leq $ flow $\leq 1$} node[below, sloped] { cost $= 0$} (17);
		\draw[->] (11) to (17);
		\draw[->] (13) to (17);
	 \draw[->] (14) -- node[above, sloped] {\footnotesize $0 \leq $ flow $\leq c_n -1$} node[below, sloped] { cost $= 0$} (17);
	\end{pgfonlayer}
\end{tikzpicture}

%% file: ms.bbl
\begin{thebibliography}{66}
\providecommand{\natexlab}[1]{#1}
\providecommand{\url}[1]{\texttt{#1}}
\expandafter\ifx\csname urlstyle\endcsname\relax
  \providecommand{\doi}[1]{doi: #1}\else
  \providecommand{\doi}{doi: \begingroup \urlstyle{rm}\Url}\fi

\bibitem[Abdulkadiro{\u{g}}lu and S{\"o}nmez(2003)]{AS03school}
Atila Abdulkadiro{\u{g}}lu and Tayfun S{\"o}nmez.
\newblock {School Choice: A Mechanism Design Approach}.
\newblock \emph{American Economic Review}, 93\penalty0 (3):\penalty0 729--747,
  2003.

\bibitem[Akrami et~al.(2022)Akrami, Chaudhury, Hoefer, Mehlhorn, Schmalhofer,
  Shahkarami, Varricchio, Vermande, and van Wijland]{ACH+22maximizing}
Hannaneh Akrami, Bhaskar~Ray Chaudhury, Martin Hoefer, Kurt Mehlhorn, Marco
  Schmalhofer, Golnoosh Shahkarami, Giovanna Varricchio, Quentin Vermande, and
  Ernest van Wijland.
\newblock {Maximizing Nash Social Welfare in 2-Value Instances}.
\newblock In \emph{Proceedings of the 36th AAAI Conference on Artificial
  Intelligence}, volume~36, pages 4760--4767, 2022.

\bibitem[Amanatidis et~al.(2023)Amanatidis, Aziz, Birmpas, Filos-Ratsikas, Li,
  Moulin, Voudouris, and Wu]{AAB+23fair}
Georgios Amanatidis, Haris Aziz, Georgios Birmpas, Aris Filos-Ratsikas, Bo~Li,
  Herv{\'e} Moulin, Alexandros~A Voudouris, and Xiaowei Wu.
\newblock {Fair Division of Indivisible Goods: Recent Progress and Open
  Questions}.
\newblock \emph{Artificial Intelligence}, page 103965, 2023.

\bibitem[Anari et~al.(2018)Anari, Mai, Gharan, and Vazirani]{AMG+18nash}
Nima Anari, Tung Mai, Shayan~Oveis Gharan, and Vijay~V Vazirani.
\newblock {Nash Social Welfare for Indivisible Items under Separable,
  Piecewise-Linear Concave Utilities}.
\newblock In \emph{Proceedings of the 29th Annual ACM-SIAM Symposium on
  Discrete Algorithms}, pages 2274--2290, 2018.

\bibitem[Babaioff et~al.(2021)Babaioff, Ezra, and Feige]{BEF21fair}
Moshe Babaioff, Tomer Ezra, and Uriel Feige.
\newblock {Fair and Truthful Mechanisms for Dichotomous Valuations}.
\newblock In \emph{Proceedings of the 35th AAAI Conference on Artificial
  Intelligence}, volume~35, pages 5119--5126, 2021.

\bibitem[Barman et~al.(2018{\natexlab{a}})Barman, Krishnamurthy, and
  Vaish]{BKV18finding}
Siddharth Barman, Sanath~Kumar Krishnamurthy, and Rohit Vaish.
\newblock {Finding Fair and Efficient Allocations}.
\newblock In \emph{Proceedings of the 2018 ACM Conference on Economics and
  Computation}, pages 557--574, 2018{\natexlab{a}}.

\bibitem[Barman et~al.(2018{\natexlab{b}})Barman, Krishnamurthy, and
  Vaish]{BKV18greedy}
Siddharth Barman, Sanath~Kumar Krishnamurthy, and Rohit Vaish.
\newblock {Greedy Algorithms for Maximizing Nash Social Welfare}.
\newblock In \emph{Proceedings of the 17th International Conference on
  Autonomous Agents and MultiAgent Systems}, pages 7--13, 2018{\natexlab{b}}.

\bibitem[Barman et~al.(2020)Barman, Bhaskar, Krishna, and
  Sundaram]{BBK+20tight}
Siddharth Barman, Umang Bhaskar, Anand Krishna, and Ranjani~G Sundaram.
\newblock {Tight Approximation Algorithms for $p$-Mean Welfare Under
  Subadditive Valuations}.
\newblock In \emph{Proceedings of the 28th Annual European Symposium on
  Algorithms}, 2020.

\bibitem[Benabbou et~al.(2021)Benabbou, Chakraborty, Igarashi, and
  Zick]{BCI+21finding}
Nawal Benabbou, Mithun Chakraborty, Ayumi Igarashi, and Yair Zick.
\newblock {Finding Fair and Efficient Allocations for Matroid Rank Valuations}.
\newblock \emph{ACM Transactions on Economics and Computation}, 9\penalty0
  (4):\penalty0 1--41, 2021.

\bibitem[Biswas and Barman(2019)]{BB19matroid}
Arpita Biswas and Siddharth Barman.
\newblock {Matroid Constrained Fair Allocation Problem}.
\newblock In \emph{Proceedings of the AAAI Conference on Artificial
  Intelligence}, volume~33, pages 9921--9922, 2019.

\bibitem[Brandt et~al.(2016)Brandt, Conitzer, Endriss, Lang, and
  Procaccia]{BCE+16handbook}
Felix Brandt, Vincent Conitzer, Ulle Endriss, J{\'e}r{\^o}me Lang, and Ariel~D
  Procaccia.
\newblock \emph{{Handbook of Computational Social Choice}}.
\newblock Cambridge University Press, 2016.

\bibitem[Bu et~al.(2023)Bu, Li, Liu, Song, and Tao]{BLL+23fair}
Xiaolin Bu, Zihao Li, Shengxin Liu, Jiaxin Song, and Biaoshuai Tao.
\newblock {Fair Division with Allocator’s Preference}.
\newblock In \emph{Proceedings of the 19th International Conference on Web and
  Internet Economics}, pages 77--94. Springer, 2023.

\bibitem[Budish(2011)]{B11combinatorial}
Eric Budish.
\newblock {The Combinatorial Assignment Problem: Approximate Competitive
  Equilibrium from Equal Incomes}.
\newblock \emph{Journal of Political Economy}, 119\penalty0 (6):\penalty0
  1061--1103, 2011.

\bibitem[Caragiannis et~al.(2019)Caragiannis, Kurokawa, Moulin, Procaccia,
  Shah, and Wang]{CKM+19unreasonable}
Ioannis Caragiannis, David Kurokawa, Herv{\'e} Moulin, Ariel~D Procaccia,
  Nisarg Shah, and Junxing Wang.
\newblock {The Unreasonable Fairness of Maximum Nash Welfare}.
\newblock \emph{ACM Transactions on Economics and Computation}, 7\penalty0
  (3):\penalty0 1--32, 2019.

\bibitem[Chaturvedi and McConnell(2017)]{CM17note}
Mmanu Chaturvedi and Ross~M McConnell.
\newblock {A Note on Finding Minimum Mean Cycle}.
\newblock \emph{Information Processing Letters}, 127:\penalty0 21--22, 2017.

\bibitem[Chaudhury et~al.(2021)Chaudhury, Garg, and Mehta]{CGM21fair}
Bhaskar~Ray Chaudhury, Jugal Garg, and Ruta Mehta.
\newblock {Fair and Efficient Allocations under Subadditive Valuations}.
\newblock In \emph{Proceedings of the 35th AAAI Conference on Artificial
  Intelligence}, volume~35, pages 5269--5276, 2021.

\bibitem[Chaudhury et~al.(2022)Chaudhury, Cheung, Garg, Garg, Hoefer, and
  Mehlhorn]{CCG+22fair}
Bhaskar~Ray Chaudhury, Yun~Kuen Cheung, Jugal Garg, Naveen Garg, Martin Hoefer,
  and Kurt Mehlhorn.
\newblock {Fair Division of Indivisible Goods For a Class of Concave
  Valuations}.
\newblock \emph{Journal of Artificial Intelligence Research}, 74:\penalty0
  111--142, 2022.

\bibitem[Cole and Gkatzelis(2018)]{CG18approximating}
Richard Cole and Vasilis Gkatzelis.
\newblock {Approximating the Nash Social Welfare with Indivisible Items}.
\newblock \emph{SIAM Journal on Computing}, 47\penalty0 (3):\penalty0
  1211--1236, 2018.

\bibitem[Cole et~al.(2017)Cole, Devanur, Gkatzelis, Jain, Mai, Vazirani, and
  Yazdanbod]{CDG+17convex}
Richard Cole, Nikhil Devanur, Vasilis Gkatzelis, Kamal Jain, Tung Mai, Vijay~V
  Vazirani, and Sadra Yazdanbod.
\newblock {Convex Program Duality, Fisher Markets, and Nash Social Welfare}.
\newblock In \emph{Proceedings of the 2017 ACM Conference on Economics and
  Computation}, pages 459--460, 2017.

\bibitem[Dobzinski et~al.(2024)Dobzinski, Li, Rubinstein, and
  Vondr{\'a}k]{DLR+24constant}
Shahar Dobzinski, Wenzheng Li, Aviad Rubinstein, and Jan Vondr{\'a}k.
\newblock {A Constant-Factor Approximation for Nash Social Welfare with
  Subadditive Valuations}.
\newblock In \emph{Proceedings of the 56th Annual ACM Symposium on Theory of
  Computing}, pages 467--478, 2024.

\bibitem[Dror et~al.(2023)Dror, Feldman, and Segal-Halevi]{DFS23fair}
Amitay Dror, Michal Feldman, and Erel Segal-Halevi.
\newblock {On Fair Division under Heterogeneous Matroid Constraints}.
\newblock \emph{Journal of Artificial Intelligence Research}, 76:\penalty0
  567--611, 2023.

\bibitem[Eisenberg and Gale(1959)]{EG59consensus}
Edmund Eisenberg and David Gale.
\newblock {Consensus of Subjective Probabilities: The Pari-Mutuel Method}.
\newblock \emph{The Annals of Mathematical Statistics}, 30\penalty0
  (1):\penalty0 165--168, 1959.

\bibitem[Feng and Li(2024)]{feng_et_al:LIPIcs.ICALP.2024.63}
Yuda Feng and Shi Li.
\newblock {A Note on Approximating Weighted Nash Social Welfare with Additive
  Valuations}.
\newblock In \emph{Procedings of the 51st International Colloquium on Automata,
  Languages, and Programming}, pages 63--1. Schloss Dagstuhl--Leibniz-Zentrum
  f{\"u}r Informatik, 2024.

\bibitem[Fitzsimmons et~al.(2024)Fitzsimmons, Viswanathan, and
  Zick]{FVZ24hardness}
Zack Fitzsimmons, Vignesh Viswanathan, and Yair Zick.
\newblock {On the Hardness of Fair Allocation under Ternary Valuations}.
\newblock \emph{arXiv preprint arXiv:2403.00943}, 2024.

\bibitem[Fleiner(2003)]{F03fixed}
Tam{\'a}s Fleiner.
\newblock {A Fixed-Point Approach to Stable Matchings and Some Applications}.
\newblock \emph{Mathematics of Operations Research}, 28\penalty0 (1):\penalty0
  103--126, 2003.

\bibitem[Fleiner and Kamiyama(2016)]{FK16matroid}
Tam{\'a}s Fleiner and Naoyuki Kamiyama.
\newblock {A Matroid Approach to Stable Matchings with Lower Quotas}.
\newblock \emph{Mathematics of Operations Research}, 41\penalty0 (2):\penalty0
  734--744, 2016.

\bibitem[Freeman et~al.(2021)Freeman, Micha, and Shah]{FMS21twosided}
Rupert Freeman, Evi Micha, and Nisarg Shah.
\newblock {Two-Sided Matching Meets Fair Division}.
\newblock In \emph{Proceedings of the 30th International Joint Conference on
  Artificial Intelligence}, pages 203--209, 2021.

\bibitem[Gale and Shapley(1962)]{GS62college}
David Gale and Lloyd~S Shapley.
\newblock {College Admissions and the Stability of Marriage}.
\newblock \emph{The American Mathematical Monthly}, 69\penalty0 (1):\penalty0
  9--15, 1962.

\bibitem[Garg and Murhekar(2021)]{GM21hardness}
Jugal Garg and Aniket Murhekar.
\newblock Computing fair and efficient allocations with few utility values.
\newblock In \emph{Proceedings of the 14th International Symposium on
  Algorithmic Game Theory}, page 345–359, 2021.

\bibitem[Garg et~al.(2018)Garg, Hoefer, and Mehlhorn]{GHM18approximating}
Jugal Garg, Martin Hoefer, and Kurt Mehlhorn.
\newblock {Approximating the Nash Social Welfare with Budget-Additive
  Valuations}.
\newblock In \emph{Proceedings of the Twenty-Ninth Annual ACM-SIAM Symposium on
  Discrete Algorithms}, pages 2326--2340, 2018.

\bibitem[Garg et~al.(2023{\natexlab{a}})Garg, Husi{\'c}, Li, V{\'e}gh, and
  Vondr{\'a}k]{GHL+23approximating}
Jugal Garg, Edin Husi{\'c}, Wenzheng Li, L{\'a}szl{\'o}~A V{\'e}gh, and Jan
  Vondr{\'a}k.
\newblock {Approximating Nash Social Welfare by Matching and Local Search}.
\newblock In \emph{Proceedings of the 55th Annual ACM Symposium on Theory of
  Computing}, pages 1298--1310, 2023{\natexlab{a}}.

\bibitem[Garg et~al.(2023{\natexlab{b}})Garg, Kulkarni, and
  Kulkarni]{GKK23approximating}
Jugal Garg, Pooja Kulkarni, and Rucha Kulkarni.
\newblock {Approximating Nash Social Welfare under Submodular Valuations
  through (un) Matchings}.
\newblock \emph{ACM Transactions on Algorithms}, 19\penalty0 (4):\penalty0
  1--25, 2023{\natexlab{b}}.

\bibitem[Garg et~al.(2024)Garg, Hoefer, and Mehlhorn]{GHM24satiation}
Jugal Garg, Martin Hoefer, and Kurt Mehlhorn.
\newblock {Satiation in Fisher Markets and Approximation of Nash Social
  Welfare}.
\newblock \emph{Mathematics of Operations Research}, 49\penalty0 (2):\penalty0
  1109--1139, 2024.

\bibitem[Goldberg and Tarjan(1989)]{GT89finding}
Andrew~V Goldberg and Robert~E Tarjan.
\newblock {Finding Minimum-Cost Circulations by Canceling Negative Cycles}.
\newblock \emph{Journal of the ACM}, 36\penalty0 (4):\penalty0 873--886, 1989.

\bibitem[Gourv{\`e}s et~al.(2014)Gourv{\`e}s, Monnot, and Tlilane]{GM14near}
Laurent Gourv{\`e}s, J{\'e}r{\^o}me Monnot, and Lydia Tlilane.
\newblock {Near Fairness in Matroids}.
\newblock In \emph{Proceedings of the 21st European Conference on Artificial
  Intelligence}, pages 393--398, 2014.

\bibitem[Gupta et~al.(2023)Gupta, Nagori, Chakraborty, Vaish, Ranu, Nadkarni,
  Dasararaju, and Chelliah]{GNC+23towards}
Anjali Gupta, Shreyans~J Nagori, Abhijnan Chakraborty, Rohit Vaish, Sayan Ranu,
  Prajit~Prashant Nadkarni, Narendra~Varma Dasararaju, and Muthusamy Chelliah.
\newblock {Towards Fair Allocation in Social Commerce Platforms}.
\newblock In \emph{Proceedings of the ACM Web Conference 2023}, pages
  3744--3754, 2023.

\bibitem[Gusfield and Irving(1989)]{GI89stable}
Dan Gusfield and Robert~W Irving.
\newblock \emph{{The Stable Marriage Problem: Structure and Algorithms}}.
\newblock MIT press, 1989.

\bibitem[Igarashi et~al.(2024)Igarashi, Kawase, Suksompong, and
  Sumita]{IKS+24fair}
Ayumi Igarashi, Yasushi Kawase, Warut Suksompong, and Hanna Sumita.
\newblock {Fair Division with Two-Sided Preferences}.
\newblock \emph{Games and Economic Behavior}, 147:\penalty0 268--287, 2024.

\bibitem[Irving et~al.(1987)Irving, Leather, and Gusfield]{ILG87efficient}
Robert~W Irving, Paul Leather, and Dan Gusfield.
\newblock {An Efficient Algorithm for the “Optimal” Stable Marriage}.
\newblock \emph{Journal of the ACM}, 34\penalty0 (3):\penalty0 532--543, 1987.

\bibitem[Jain and Vaish(2024)]{JV24maximizing}
Pallavi Jain and Rohit Vaish.
\newblock {Maximizing Nash Social Welfare under Two-Sided Preferences}.
\newblock In \emph{Proceedings of the 38th AAAI Conference on Artificial
  Intelligence}, volume~38, pages 9798--9806, 2024.

\bibitem[Kamiyama(2015)]{K15stable}
Naoyuki Kamiyama.
\newblock {Stable Matchings with Ties, Master Preference Lists, and Matroid
  Constraints}.
\newblock In \emph{Proceedings of the 8th International Symposium on
  Algorithmic Game Theory}, pages 3--14. Springer, 2015.

\bibitem[Kamiyama(2017)]{K17popular}
Naoyuki Kamiyama.
\newblock {Popular Matchings with Ties and Matroid Constraints}.
\newblock \emph{SIAM Journal on Discrete Mathematics}, 31\penalty0
  (3):\penalty0 1801--1819, 2017.

\bibitem[Kaneko and Nakamura(1979)]{KN79nash}
Mamoru Kaneko and Kenjiro Nakamura.
\newblock {The Nash Social Welfare Function}.
\newblock \emph{Econometrica: Journal of the Econometric Society}, pages
  423--435, 1979.

\bibitem[Karp(1978)]{K78characterization}
Richard~M Karp.
\newblock {A Characterization of the Minimum Cycle Mean in a Digraph}.
\newblock \emph{Discrete Mathematics}, 23\penalty0 (3):\penalty0 309--311,
  1978.

\bibitem[Khot et~al.(2008)Khot, Lipton, Markakis, and
  Mehta]{KLM+08inapproximability}
Subhash Khot, Richard~J Lipton, Evangelos Markakis, and Aranyak Mehta.
\newblock {Inapproximability Results for Combinatorial Auctions with Submodular
  Utility Functions}.
\newblock \emph{Algorithmica}, 52:\penalty0 3--18, 2008.

\bibitem[Knuth(1997)]{K97stable}
Donald~Ervin Knuth.
\newblock \emph{{Stable Marriage and its Relation to Other Combinatorial
  Problems: An Introduction to the Mathematical Analysis of Algorithms}},
  volume~10.
\newblock American Mathematical Soc., 1997.

\bibitem[Lee(2017)]{L17apx}
Euiwoong Lee.
\newblock {APX-Hardness of Maximizing Nash Social Welfare with Indivisible
  Items}.
\newblock \emph{Information Processing Letters}, 122:\penalty0 17--20, 2017.

\bibitem[Manlove(2013)]{M13algorithmics}
David Manlove.
\newblock \emph{{Algorithmics of Matching under Preferences}}, volume~2.
\newblock World Scientific, 2013.

\bibitem[Markakis(2017)]{M17approximation}
Evangelos Markakis.
\newblock {Approximation Algorithms and Hardness Results for Fair Division with
  Indivisible Goods}.
\newblock \emph{Trends in Computational Social Choice}, pages 231--247, 2017.

\bibitem[McVitie and Wilson(1971)]{MW71stable}
David~G McVitie and Leslie~B Wilson.
\newblock {The Stable Marriage Problem}.
\newblock \emph{Communications of the ACM}, 14\penalty0 (7):\penalty0 486--490,
  1971.

\bibitem[Moulin(2004)]{M04fair}
Herv{\'e} Moulin.
\newblock \emph{{Fair Division and Collective Welfare}}.
\newblock MIT press, 2004.

\bibitem[Narang et~al.(2022)Narang, Biswas, and Narahari]{NBN22achieving}
Shivika Narang, Arpita Biswas, and Yadati Narahari.
\newblock {On Achieving Leximin Fairness and Stability in Many-to-One
  Matchings}.
\newblock In \emph{Proceedings of the 21st International Conference on
  Autonomous Agents and Multiagent Systems}, pages 1705--1707, 2022.

\bibitem[Nash~Jr(1950)]{N50bargaining}
John~F Nash~Jr.
\newblock {The Bargaining Problem}.
\newblock \emph{Econometrica: Journal of the Econometric Society}, pages
  155--162, 1950.

\bibitem[Nguyen et~al.(2014)Nguyen, Nguyen, Roos, and
  Rothe]{NNR+14computational}
Nhan-Tam Nguyen, Trung~Thanh Nguyen, Magnus Roos, and J{\"o}rg Rothe.
\newblock {Computational Complexity and Approximability of Social Welfare
  Optimization in Multiagent Resource Allocation}.
\newblock \emph{Autonomous Agents and Multi-Agent Systems}, 28\penalty0
  (2):\penalty0 256--289, 2014.

\bibitem[Nisan et~al.(2007)Nisan, Roughgarden, Tardos, and Vazirani]{NRT+07agt}
Noam Nisan, Tim Roughgarden, \'Eva Tardos, and Vijay~V. Vazirani.
\newblock \emph{Algorithmic Game Theory}.
\newblock Cambridge University Press, New York, NY, USA, 2007.

\bibitem[Roth and Peranson(1999)]{RP99redesign}
Alvin~E Roth and Elliott Peranson.
\newblock {The Redesign of the Matching Market for American Physicians: Some
  Engineering Aspects of Economic Design}.
\newblock \emph{American Economic Review}, 89\penalty0 (4):\penalty0 748--780,
  1999.

\bibitem[Roth and Sotomayor(1992)]{RS92two}
Alvin~E Roth and Marilda Sotomayor.
\newblock {Two-Sided Matching}.
\newblock \emph{Handbook of Game Theory with Economic Applications},
  1:\penalty0 485--541, 1992.

\bibitem[Roth et~al.(2004)Roth, S{\"o}nmez, and {\"U}nver]{RSU04kidney}
Alvin~E Roth, Tayfun S{\"o}nmez, and M~Utku {\"U}nver.
\newblock {Kidney Exchange}.
\newblock \emph{The Quarterly Journal of Economics}, 119\penalty0 (2):\penalty0
  457--488, 2004.

\bibitem[Sethuraman et~al.(2006)Sethuraman, Teo, and Qian]{STQ06many}
Jay Sethuraman, Chung-Piaw Teo, and Liwen Qian.
\newblock {Many-To-One Stable Matching: Geometry and Fairness}.
\newblock \emph{Mathematics of Operations Research}, 31\penalty0 (3):\penalty0
  581--596, 2006.

\bibitem[Shoshan et~al.(2023)Shoshan, Hazon, and Segal-Halevi]{SHS23efficient}
Hila Shoshan, Noam Hazon, and Erel Segal-Halevi.
\newblock {Efficient Nearly-Fair Division with Capacity Constraints}.
\newblock In \emph{Proceedings of the 2023 International Conference on
  Autonomous Agents and Multiagent Systems}, pages 206--214, 2023.

\bibitem[Suksompong(2021)]{S21constraints}
Warut Suksompong.
\newblock {Constraints in Fair Division}.
\newblock \emph{ACM SIGecom Exchanges}, 19\penalty0 (2):\penalty0 46--61, 2021.

\bibitem[Teo and Sethuraman(1998)]{TS98geometry}
Chung-Piaw Teo and Jay Sethuraman.
\newblock {The Geometry of Fractional Stable Matchings and its Applications}.
\newblock \emph{Mathematics of Operations Research}, 23\penalty0 (4):\penalty0
  874--891, 1998.

\bibitem[Tomita and Yokoyama(2024)]{TY24fair}
Yoji Tomita and Tomohiki Yokoyama.
\newblock {Fair Reciprocal Recommendation in Matching Markets}.
\newblock \emph{arXiv preprint arXiv:2409.00720}, 2024.

\bibitem[Viswanathan and Zick(2023)]{VZ23general}
Vignesh Viswanathan and Yair Zick.
\newblock {A General Framework for Fair Allocation under Matroid Rank
  Valuations}.
\newblock In \emph{Proceedings of the 24th ACM Conference on Economics and
  Computation}, pages 1129--1152, 2023.

\bibitem[Vondr{\'a}k(2008)]{V08optimal}
Jan Vondr{\'a}k.
\newblock {Optimal Approximation for the Submodular Welfare Problem in the
  Value Oracle Model}.
\newblock In \emph{Proceedings of the 40th Annual ACM Symposium on Theory of
  Computing}, pages 67--74, 2008.

\bibitem[Wu et~al.(2021)Wu, Li, and Gan]{WLG21budget}
Xiaowei Wu, Bo~Li, and Jiarui Gan.
\newblock {Budget-Feasible Maximum Nash Social Welfare is Almost Envy-free}.
\newblock In \emph{Proceedings of the 30th International Joint Conference on
  Artificial Intelligence}, pages 465--471, 2021.

\end{thebibliography}
